\documentclass[%
 reprint,
superscriptaddress,
 amsmath,amssymb,
 aps,pre
]{revtex4-2}

\usepackage{graphicx}%
\usepackage{dcolumn}%
\usepackage{bm}%
\usepackage{color}

\begin{document}

\title{Physical limits on galvanotaxis}%
\author{Ifunanya Nwogbaga}
\thanks{These two authors contributed equally}
\affiliation{Thomas C. Jenkins Department of Biophysics, Johns Hopkins University}%
\author{A Hyun Kim}
\thanks{These two authors contributed equally}
\affiliation{Thomas C. Jenkins Department of Biophysics, Johns Hopkins University}%
\author{Brian A. Camley}
\affiliation{Thomas C. Jenkins Department of Biophysics, Johns Hopkins University}%
\affiliation{William H. Miller III Department of Physics \& Astronomy, Johns Hopkins University}

\begin{abstract}
Eukaryotic cells can polarize and migrate in response to electric fields via ``galvanotaxis," which aids wound healing. Experimental evidence suggests cells sense electric fields via molecules on the cell's surface redistributing via electrophoresis and electroosmosis, though the sensing species has not yet been conclusively identified. We develop a model that links sensor redistribution and galvanotaxis using maximum likelihood estimation. 
Our model predicts a single universal curve for how galvanotactic directionality depends on field strength. We can collapse measurements of galvanotaxis in keratocytes, neural crest cells, and granulocytes to this curve, suggesting that stochasticity due to the finite number of sensors may limit galvanotactic accuracy. We find cells can achieve experimentally observed directionalities with either a few ($\sim100$) highly-polarized sensors, or many ($\sim10^4$) sensors with a $\sim6-10\%$ change in concentration across the cell. We also identify additional signatures of galvanotaxis via sensor redistribution, including the presence of a tradeoff between accuracy and variance in cells being controlled by rapidly switching fields. Our approach shows how the physics of noise at the molecular scale can limit cell-scale galvanotaxis, providing important constraints on sensor properties, and allowing for new tests to determine the specific molecules underlying galvanotaxis.
\end{abstract} 

\maketitle

Eukaryotic cells will migrate  singly or in groups in response to an applied electrical field -- a process called ``galvanotaxis" or ``electrotaxis" \cite{sengupta2021principles,sun2013keratocyte,zajdel2020scheepdog,shim2021overriding,lalli2015collective,dawson2022cell,sun2020pi3k,sun2023electric,zhang2022propagation}. Response to electric fields helps give directionality to wound healing \cite{kennard2020osmolarity} and immune response \cite{sun2019infection}, overriding other guidance cues \cite{zhao2009electrical}. While galvanotactic responses have been  measured for decades, understanding of the mechanisms of galvanotaxis pale in comparison to chemotaxis,  where cells respond to chemical gradients \cite{sengupta2021principles}. The current best-supported theory is that galvanotaxis arises because of migration of a sensor species on the surface of the cell in response to the electric field \cite{allen2013electrophoresis,kobylkevich2018reversing,sarkar2019electromigration} (ciliated cells may have an alternate behavior \cite{ogawa2006physical}).  Single-cell galvanotaxis has been modeled phenomenologically \cite{prescott2021quantifying,nwogbaga2022coupling,gruler1991neural,franke1990galvanotaxis,schienbein1993langevin} but these models do not connect galvanotaxis to sensor rearrangements. 

There is strong evidence that both eukaryotic cells and bacteria can sense chemicals at nearly the limits imposed on them by basic physical and statistical principles \cite{fuller2010external,berg1977physics,mattingly2021escherichia,andrews2007information,segota2013high,song2006dictyostelium,van2007biased},  and broader interest in finding fundamental physical bounds for accuracy \cite{ipina2022collective,fancher2017fundamental,fancher2020precision,vennettilli2021precision,bialek2005physical,mora2019physical,badvaram2022physical}. Here, we ask: how precisely can a cell sense the direction of an applied electrical field? Our strategy will be to write a model for the probability of observing a sensor configuration in the field, and then determine how the cell can estimate the field angle by choosing an estimated direction $\hat{\psi}$ that maximizes the likelihood of observing this configuration. This approach builds off past maximum-likelihood estimation (MLE) results that established optimal ways to sense chemical concentrations and gradients \cite{endres2009maximum,hu2010physical,hu2011geometry,hopkins2020chemotaxis,camley2017cell,mora2019physical}. We derive results bounding the best possible estimation cells can make of the field orientation given the unavoidable randomness in sensor positions. 

\begin{figure*}[htb]
    \centering
    \includegraphics[width=\textwidth]{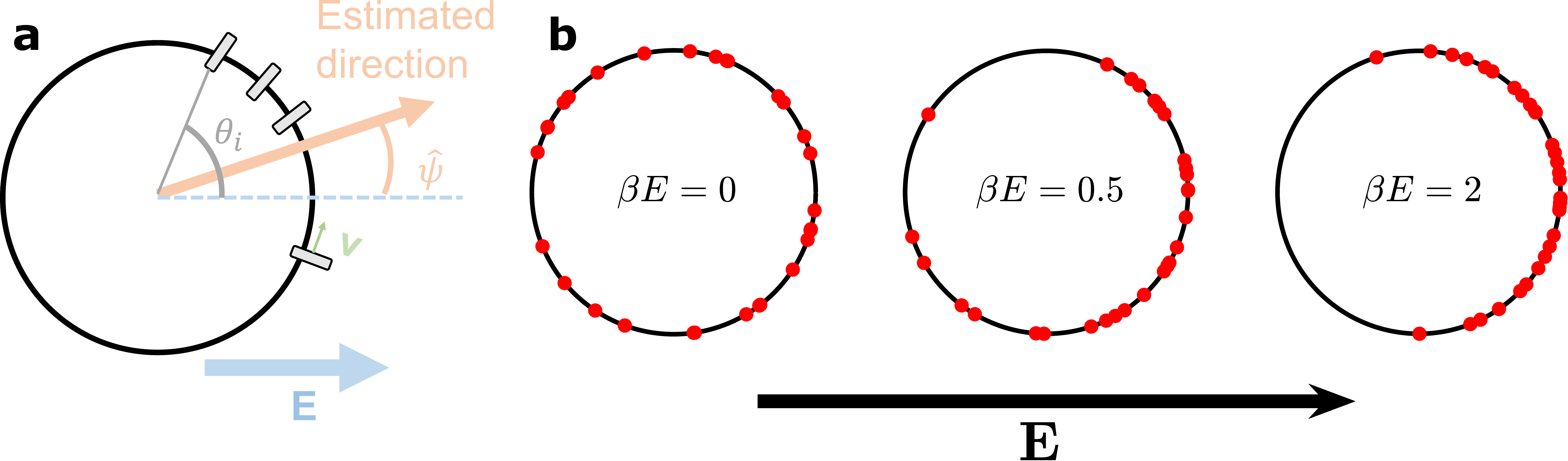}
    \caption{Illustration of sensors on surface of cell and schematic of sensor distributions. {\bf a:} The cell uses the stochastic locations of  sensors $(\theta_1,\theta_2,\cdots,\theta_N)$ to make a noisy estimate $\hat{\psi}$ of the field's true direction $\psi$. Sensors travel with velocity $\mathbf{v}$ proportional to the component of the electric field tangential to the membrane. {\bf b:} Stochastic simulation of $N=30$ sensor positions for different field strengths, generated by drawing sensor positions independently from the distribution $p(\theta) \sim \exp({\kappa \cos (\theta-\psi))}$.}
    \label{fig:illus}
\end{figure*}

\section*{Sensor migration model}
We apply a simple model, assuming that a receptor -- or other charged molecule -- migrates on the surface of the cell in response to an electric field. We call this molecule a {\it sensor}. The prevailing consensus is that sensor redistribution via electrophoresis and electroosmosis is necessary for a galvanotactic response \cite{allen2013electrophoresis,kobylkevich2018reversing,sarkar2019electromigration,mclaughlin1981role,feder1994redistribution}. Other proposed mechanisms such as 
asymmetric opening of voltage gated ion channels %
are not supported by evidence on, e.g. changing the viscosity of the medium surrounding the cells \cite{allen2013electrophoresis,kobylkevich2018reversing}. %

We assume the sensor species travels along the cell membrane with a velocity $\mathbf{v}$ that is proportional to the component of electric field in the membrane's tangent plane \cite{mclaughlin1981role}. For many different assumptions about the cell and membrane properties, this leads to an ``effective mobility'' $\mu$ where the sensor velocity is $v_\parallel = \mu E_\parallel$, i.e.  $\mathbf{v} = \mu \left[\mathbf{E} -  (\mathbf{E}\cdot \hat{\mathbf{n}}) \hat{\mathbf{n}}\right]$ where $\hat{\mathbf{n}}$ is a local normal to the surface of the cell  and $\mathbf{E}$ is the {\it applied} electric field. The parameter $\mu$, which can be positive or negative, includes the effect of electrophoresis and electroosmotic flow  and the effect of the cell on the electric field (see Appendix \ref{app:efield_details}) and can be estimated from microscopic properties of the sensor and environment \cite{mclaughlin1981role,sarkar2019electromigration,kobylkevich2018reversing}; we will interpret $\mu$ as purely a phenomenological value. 

The steady-state probability distribution of sensor locations arises from competition between sensor advection due to the field and diffusion tending to spread the sensors out. The probability flux from  sensor migration and diffusion on the cell surface is $\mathbf{J} = \mathbf{v} p - D\nabla_S p$ where $p$ is the probability density on the surface, $D$ is the sensor diffusion coefficient, and $\nabla_S$ the surface gradient. The continuity equation is then $\frac{\partial p}{\partial t} = -\nabla_S \cdot \mathbf{J}$. We initially consider a two-dimensional model of a cell as a circle. %
Positions on the cell surface are parameterized by an angle $\theta$, and $\nabla_S \to \frac{1}{R} \partial_\theta$. For our circle, the velocity is $\mathbf{v} = \mu \left[\mathbf{E} -  (\mathbf{E}\cdot \hat{\mathbf{n}}) \hat{\mathbf{n}}\right] = \mu (\mathbf{E}\cdot\hat{\bm{\theta}}) \hat{\bm{\theta}}$, where $\hat{\bm{\theta}} = (-\sin\theta,\cos\theta)$ is the local tangent. Our continuity equation becomes, if the electrical field is in the $\psi$ direction,
\begin{equation}
    \frac{\partial}{\partial t} p(\theta,t) = -\frac{1}{R} \frac{\partial}{\partial \theta} \left[ -\mu E \sin (\theta-\psi) p(\theta,t) \right] + \frac{D}{R^2} \frac{\partial^2 }{\partial \theta^2} p(\theta,t). \label{eq:continuity_general}
\end{equation}
The steady-state solution of this equation is
\begin{equation}
        p(\theta) = Z^{-1} e^{\kappa \cos \left(\theta-\psi\right)},
    \label{eq:Smoluchowski_ss}
\end{equation} 
where $\psi$ is the field's orientation relative to the $x$ axis and $\kappa = \mu E R / D$, given the field strength $E$, cell radius $R$, and diffusion constant $D$. $Z$ is a normalizing factor. $p(\theta)$ is a von Mises distribution -- a generalization of a Gaussian distribution to a periodic domain \cite{mardia2000directional}.  $\kappa$ can be interpreted as a P$\acute{\mathrm{e}}$clet number \cite{leal2007advanced}, the ratio between the timescales of diffusive spreading versus advective transport via the electric field. Increasing $\kappa$, e.g. by making the field larger, or the diffusion coefficient of the sensor smaller, means sensors are more localized (more front/back polarization). In the limit of $\kappa \gg 1$, Eq. \ref{eq:Smoluchowski_ss} becomes a Gaussian with variance $1/\kappa$. We will often think about how cell responses depend on electric field, so we also define $\beta = \mu R / D$ so $\kappa = \beta E$. $1/\beta$ is the electric field at which the cathode-anode ratio $p(\psi)/p(\psi+\pi)$ reaches $e^2\approx 7.4$. Sensor positions arising from this $p(\theta)$ as $\beta E$ is increased are plotted in Fig. \ref{fig:illus}b. We also solve for $p$ in a spherical geometry (Appendix \ref{app:sphere}).

\section*{Estimating field directions using maximum likelihood}
We assume the cell chooses an orientation $\psi$ that maximizes the likelihood  $\mathcal{L}(\psi,E;\boldsymbol{\theta}) = p(\boldsymbol\theta | \psi, E)$ given the $N$ observed sensor locations $\bm{\theta} = (\theta_1,\theta_2,\theta_3,\cdots,\theta_N)$. If the sensor positions are independent, then  $\mathcal{L}(\psi,E;\bm{\theta}) = \prod_{i=1}^{N} p(\theta_i | \psi, E)$, or
\begin{align}
    \ln \mathcal{L}(\psi,E;\bm{\theta}) &= - N \ln Z + \kappa \sum_{i=1}^N \cos (\theta_i - \psi) \\
    &= - N \ln Z + \beta \mathbf{E}\cdot\bm{\rho},
    \label{eq:log_like_circle}
\end{align}
where $\bm{\rho} = \sum_i (\cos \theta_i,\sin \theta_i)$ is the sum of sensor positions  and the field $\mathbf{E}=E(\cos\psi,\sin\psi)$. 
Given $\bm{\theta}$, the direction $\psi$ that maximizes $\ln \mathcal{L}$ is the maximum likelihood estimator of the field direction $\psi$; we call this estimator $\hat{\psi}$. We can see directly from Eq. \eqref{eq:log_like_circle} that log-likelihood is maximized if $\mathbf{E}$ \textit{is in the direction of the sum of sensor locations $\boldsymbol{\rho}$}, i.e. 
\begin{equation}
    \label{eq:receptorsum}
    \tan \hat{\psi} = \dfrac{\sum_i \sin\theta_i}{\sum_i \cos \theta_i},
\end{equation}
if $\beta > 0$ (Appendix \ref{app:MLE}). An analogous result can be derived for a sphere (Appendix \ref{app:sphere}).

The precision with which a cell can sense the direction $\psi$ is limited by the Fisher information $\mathcal{I}(\psi) = -\left\langle \frac{\partial^2 \ln \mathcal{L}}{\partial \psi^2}\right\rangle$ \cite{kay1993fundamentals}, which can be computed as $\mathcal{I}(\psi) = \kappa \sum_{i=1}^N \langle \cos(\theta_i - \psi) \rangle$. We find
\begin{equation}
    \mathcal{I}(\psi) = N \kappa \dfrac{I_1(\kappa)}{I_0(\kappa)} \; \; \; \; \textrm{(circle)}, \label{eq:fisher_circle}
\end{equation}
where $I_\nu(x)$ is a modified Bessel function of the first kind. This is also a known result for independent von Mises measurements \cite{mardia2000directional}. We can also compute the Fisher information and maximum likelihood estimators for sensors on a sphere,  assuming the cell only estimates the azimuthal field angle, using its substrate to constrain the field's plane (Appendix \ref{app:sphere}).
\begin{equation}
    \mathcal{I}(\psi) = N (\kappa \coth \kappa -1) \; \; \; \; \textrm{(sphere)}. \label{eq:sphere}
\end{equation}
The accuracy of unbiased estimators $\hat{\psi}$ of a parameter $\psi$ are limited by the Cramer-Rao bound \cite{kay1993fundamentals},
\begin{equation}
    \langle (\hat{\psi} - \psi)^2 \rangle \ge \mathcal{I}(\psi)^{-1} \; \textrm{(Cramer-Rao)}. \label{eq:cramer}
\end{equation}
However, this bound can be incorrect when estimating a direction like $\psi$ when $\mathcal{I}(\psi)$ is small.  The Cramer-Rao definition of an unbiased estimator is that $\langle \hat{\psi} \rangle = \psi$. Angles, though, may vary by factors of $2\pi$ -- e.g. an estimator with $\langle\hat{\psi}\rangle = \psi + 2\pi$ would be biased in the usual definition, but unbiased in a circular sense, requiring generalizations of Cramer-Rao
\cite{mardia2000directional}. We find a bound for circularly-defined variables (Appendix \ref{app:periodic_cramer}):
\begin{equation}
    \langle \cos (\hat{\psi} - \psi) \rangle \le \sqrt{\frac{\mathcal{I}(\psi)}{1+\mathcal{I}(\psi)}} \;  \textrm{(periodic Cramer-Rao)}. \label{eq:periodic_bound}
\end{equation}

\begin{figure}
    \centering
    \includegraphics[width=\columnwidth]{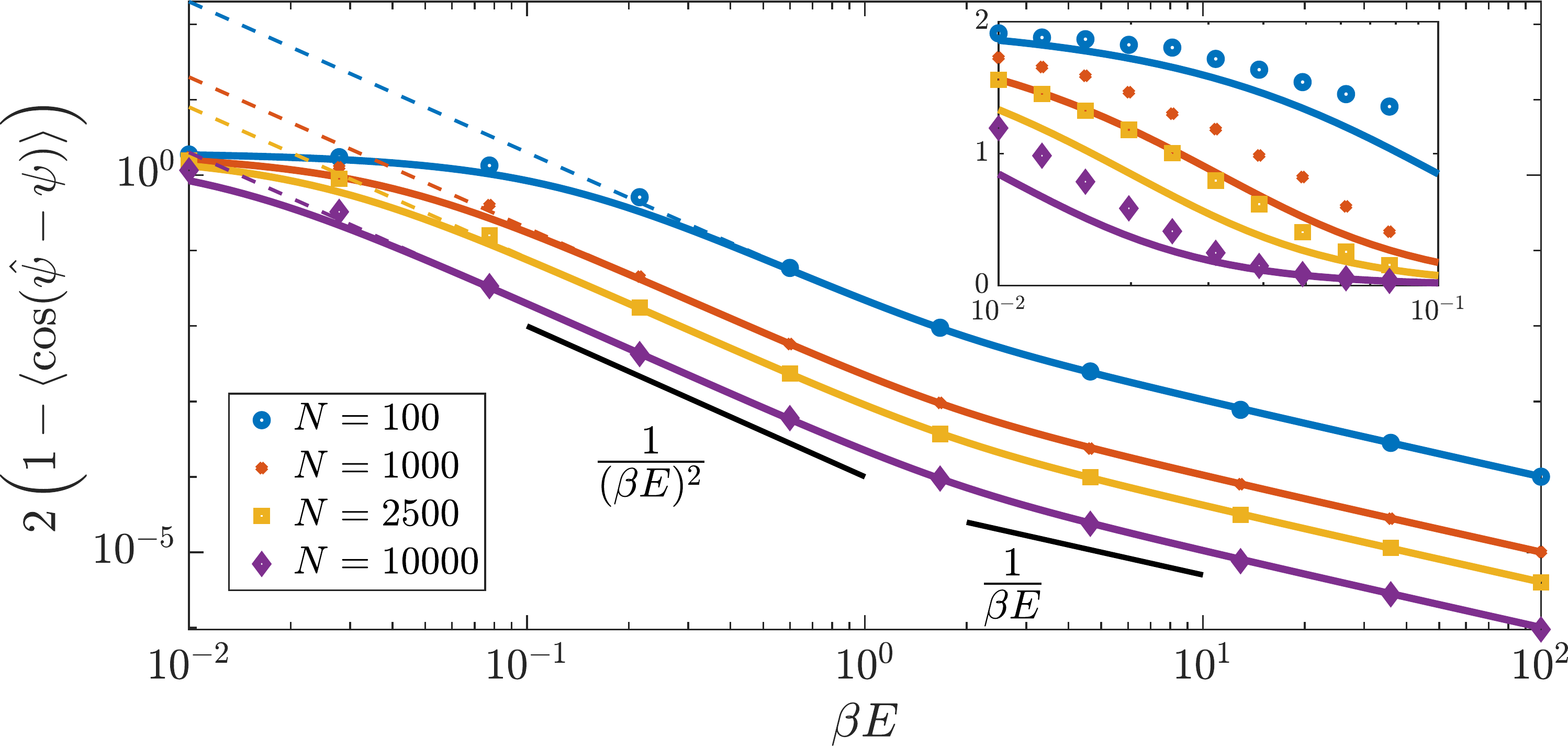}
    \caption{Accuracy of galvanotaxis  plotted as circular variance $V = 2(1-\langle \cos(\hat{\psi}-\psi)\rangle)$. Accuracy can be increased by increasing the number of sensors $N$ or making the sensors more polarized across the cell (increasing $\beta E$). Solid lines are the periodic bound (Eq. \eqref{eq:periodic_bound}), dashed lines are the normal Cramer-Rao bound (Eq. \eqref{eq:cramer}), symbols are stochastic simulation.  $d=2$ shown here; see Appendix \ref{app:sphere} for sphere.  Inset shows zoomed in region in linear scale:  MLE variance systematically exceeds Eq. \eqref{eq:periodic_bound}.}
    \label{fig:circular_variance_simple}
\end{figure}

Galvanotaxing cells are more accurate in their field sensing when the field strength $E$ is increased, or if there are more sensors (larger $N$), or if the sensors are more susceptible to the field (larger $\beta$) (Fig. \ref{fig:circular_variance_simple}). We plot the circular equivalent of the variance $V \equiv 2(1-\langle \cos(\hat{\psi}-\psi)\rangle)$; when $|\hat{\psi}-\psi| \ll 2\pi$, $V$ reduces to the ordinary variance, as can be seen via Taylor expansion. We show the periodic Cramer-Rao bound, Eq. \eqref{eq:periodic_bound} (solid lines), and $V$ computed by stochastic simulation (symbols). 

To verify our results on the the variance of the maximum-likelihood estimator shown in Fig. \ref{fig:circular_variance_simple}, we use stochastic simulation. To do this, we draw each sensor position independently from the distribution $p(\theta) \sim e^{\kappa \cos(\theta-\psi)}$. Then, for each sensor configuration of $N$ sensors, we determine the maximum-likelihood direction $\hat{\psi}$ by computing the sum of sensor positions $\boldsymbol\rho$. We then compute the average $V = 2(1-\langle \cos ( \hat{\psi}-\psi) \rangle)$ over 5000 generated sensor configurations.

The simulated circular variance $V$ agrees well with the ordinary Cramer-Rao bound at asymptotically large fields (dashed line, Eq. \eqref{eq:cramer}), with $V \approx 1/\mathcal{I}(\psi)$. Given less information about field direction, $1/\mathcal{I}(\psi)$ exceeds the maximum possible $V = 2$, and the periodic Cramer-Rao bound is a better measure of the simulated variance. The bound in Eq. \eqref{eq:periodic_bound} is not ``tight'' -- the maximum likelihood estimator  does not achieve the bound at $\beta E \ll 1$
(Fig. \ref{fig:circular_variance_simple} inset), though it is efficient at large $E$. This may occur because the bound in Eq. \eqref{eq:periodic_bound} could be improved, or MLE is not efficient for this problem \cite{kay1993fundamentals}.

\section*{Universal curve describes sensing across cell types}
In the limit of weakly polarized cells ($\kappa \lesssim 1$), which we think is likely experimentally relevant (see next section), we can make a particularly simple prediction. In this limit, $\mathcal{I}(\psi) \approx N \kappa^2 / d$, with $d = 2, 3$ for circle and sphere, respectively.  If cells perform near their ideal abilities (Eq. \eqref{eq:periodic_bound}), the dependence of directionality on  electric field will then be 
\begin{align}\langle \cos(\hat{\psi}-\psi)\rangle &= \sqrt{\dfrac{N \kappa^2 d^{-1}}{1 + N \kappa^2 d^{-1}}} \\&\equiv \sqrt{\dfrac{\gamma^2 E^2}{1 + \gamma^2 E^2}} \label{eq:simple_gamma_curve},
\end{align}
where we have collected all the unknowns into $\gamma^2 \equiv N \beta^2/d$, a single remaining fit parameter. $1/\gamma$ is the field at which the best possible directionality is $1/\sqrt{2}\approx0.7$. Eq. \eqref{eq:simple_gamma_curve} is also appropriate if there are multiple sensor types, though with a generalized $\gamma$ (Appendix \ref{app:types}).

\begin{figure}[htb]
    \centering
    \includegraphics[width=\linewidth]{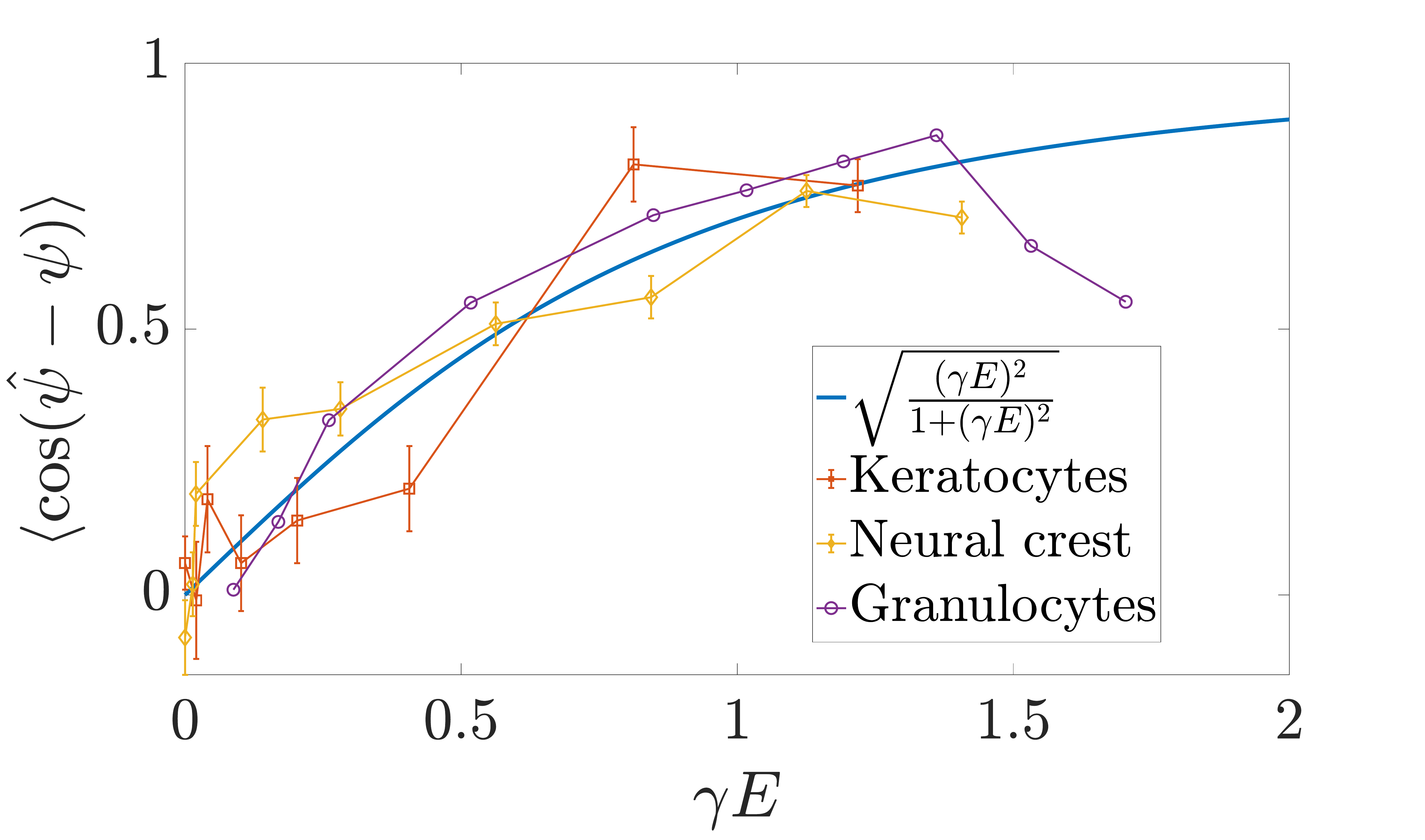}
    \caption{ Electric field dependence of experiments roughly collapse to predicted curve with one fit parameter $\gamma$. $\gamma-$values are $2\times 10^{-3}$ mm/mV, $2.8 \times 10^{-3}$ mm/mV, and $1.7 \times 10^{-3}$ mm/mV for keratocytes \cite{sun2013keratocyte}, neural crest (Fig. 1 of \cite{gruler1991neural}), and granulocytes \cite{franke1990galvanotaxis}, respectively. Granulocyte error bars are unknown.}
    \label{fig:collapse}
\end{figure}

We test this prediction in Fig. \ref{fig:collapse}, which shows three experimental measurements of galvanotaxis in different cell types: keratocytes \cite{sun2013keratocyte}, neural crest \cite{gruler1991neural}, and granulocytes \cite{franke1990galvanotaxis}. These experiments observe the cosine of the angle of cell velocity relative to the electric field -- we write  this ``directionality'' as $\langle \cos (\hat{\psi}-\psi) \rangle$, which assumes that the cell's velocity is its best estimate of the field direction. (We address generalizations of this assumption in Appendix \ref{app:downstream}.)
The experimental measurements can be reasonably collapsed onto our prediction of Eq. \eqref{eq:simple_gamma_curve}, fitting  $\gamma$ for each cell type. We see some deviations from the model at large electric fields (neural crest, granulocytes) and small electric fields (neural crest). Large-field deviations may arise from  heating and membrane damage \cite{allen2013electrophoresis,pliquett2007high}. Alternate fits  including additional sources of error like downstream noise in cell motility  \cite{fuller2010external} are discussed in Appendix \ref{app:downstream}, along with potential reasons for the neural crest deviation. A similar collapse of neural crest and granulocyte data was discovered by \cite{gruler1991neural}, though without identifying Eq. \eqref{eq:simple_gamma_curve}. 

{The data in Fig. \ref{fig:collapse} provides a tantalizing suggestion that fluctuations in sensor positions may limit galvanotactic accuracy at low field strengths, akin to earlier results on information limitation in chemotaxis \cite{fuller2010external,andrews2007information}. However, a full test of this idea would require a more confident identification of a putative sensor -- fixing $\beta$ and $N$ -- or more precise measurements to test the shape of the curve in Fig. \ref{fig:collapse}. Currently, it is impossible to rule out a model in which cells can sense the electric field near-perfectly -- but choose to respond to it in a noisy way (Appendix \ref{app:downstream}).}

\section*{Model constrains sensor properties}
What does the fit values of $\gamma = \sqrt{N \beta^2/d}$ from Fig. \ref{fig:collapse} tell us about the sensor? Are these parameters plausible? The key unknowns are $\beta$ and the number of sensors $N$. Without an established identity of the sensor, estimating $N$ is difficult. Receptors like EGFR which may play a role in galvanotaxis \cite{fang1999epidermal} can have expression levels of $\sim 5\times10^4$ receptors/cell \cite{macdonald2008heterogeneity}, but other (larger) putative sensors \cite{lin2017lipid} might be fewer in number. We can find $\beta$ for a molecule given its distribution in an electric field $p(\theta)$. Recent work measured redistribution of fluorescent tdTomato-GPI in an electric field, finding a cathode/anode fluorescence ratio of roughly 2.0 at $E = 1000$ mV/mm \cite{sarkar2019electromigration}. This ratio's dependence on field strength was consistent with  electromigration  \cite{mclaughlin1981role,kobylkevich2018reversing,sarkar2019electromigration}.  %
In our electromigration model the ratio between cathodal-side and anodal-side probability density is $e^{2\kappa}$. The data on tdTomato-GPI then shows $e^{2\kappa} = 2$, or  $\kappa \approx 0.35$ at $E = 1000$ mV/mm, or $\beta \approx (0.35) / (1000~\textrm{mV/mm}) \approx 3.5 \times 10^{-4}$ mm/mV. This supports our assumption that $\kappa \lesssim 1$ in the previous section: 1000 mV/mm is a strong field, as some cells migrate directionally in fields of $\sim$ 10 mV/mm \cite{gruler1991neural}. Fields {\it in vivo} have been measured at $40-200$ mV/mm (mammalian wounds \cite{pullar2005cyclic}) and $27-40$ mV/mm ({\it Xenopus} embryonic development \cite{hotary1994endogenous}).

Because the only relevant fitting parameter is $\gamma^2 \equiv N \beta^2 / d$, we cannot separately determine $N$ and $\beta$ -- i.e. we cannot tell how much of the cell's accuracy is driven by having a large number of sensors vs sensors that are highly polarized. If we use $\beta \approx 3.5 \times 10^{-4}$ mm/mV appropriate to tdTomato-GPI experiments on CHO cells and $d = 3$,   all three cell types must have roughly $70-200$ sensors. By contrast, if we assume that the field sensor is expressed at a level similar to typical chemoattractant receptors, guessing $N \approx 10^4$, then we find $\beta \approx (3-5) \times 10^{-5}$ mm/mV for the cell types studied here. This would correspond to a cathode/anode fluorescence ratio of $e^{2\beta E} \approx 1.06-1.10$ at $E = 1000 $ mV/mm. This implies that {\it the sensor need not be strongly polarized, even at large fields} -- similar to the observation that as few as a hundred bound receptors' difference between front and back may lead to chemotactic migration \cite{song2006dictyostelium}.

\begin{figure}[htb]
    \centering
    \includegraphics[width=\linewidth]{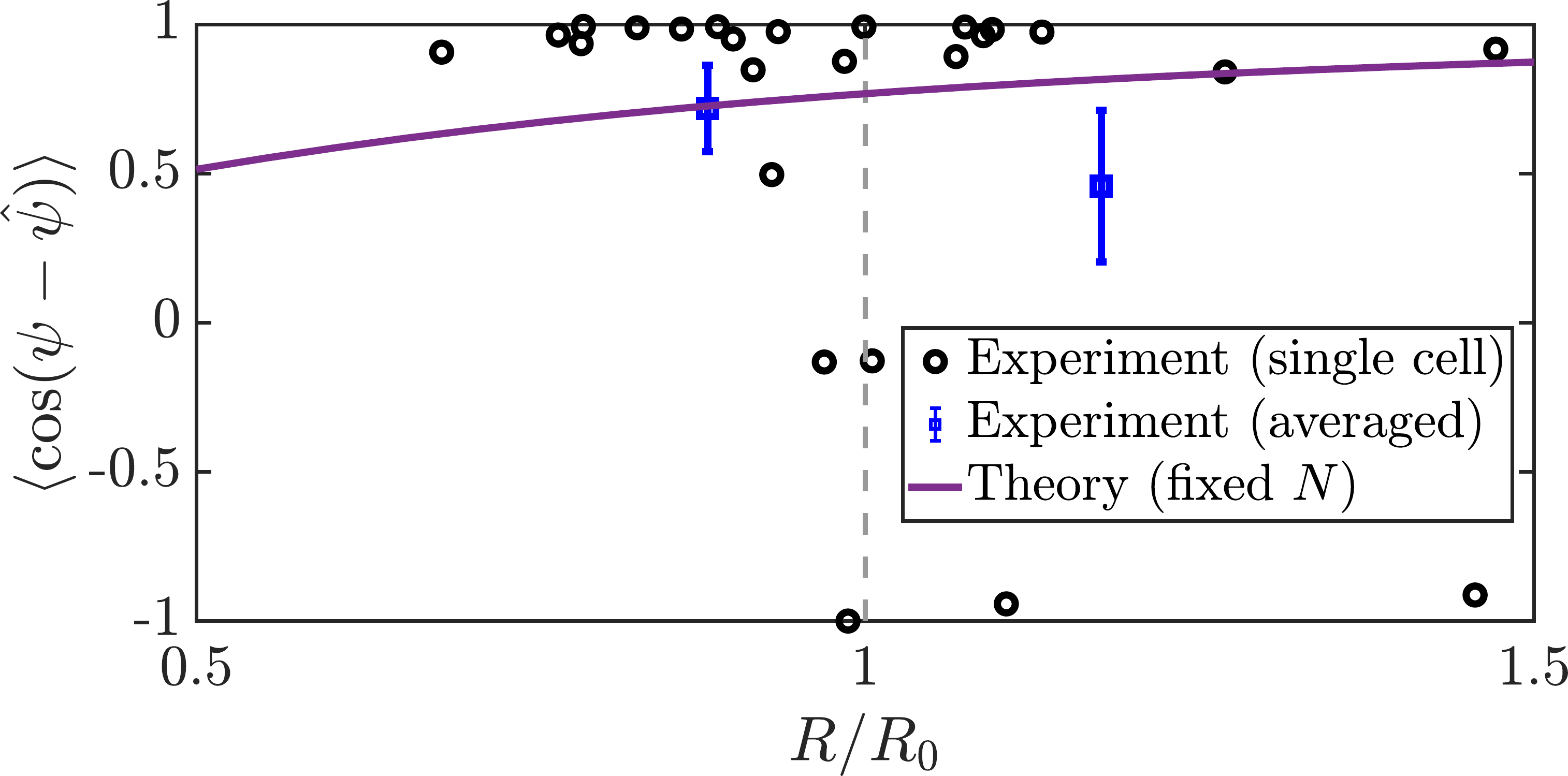}
    \caption{ Directionality varies with cell radius in the model; experimental data are not as clear. Lines are the bound (Eq. \eqref{eq:periodic_bound} with Eq. \eqref{eq:fisher_circle}), symbols are experimental data on keratocytes from \cite{sun2013keratocyte}. Error bars are standard error. $\gamma = 0.002$ mm/mV is the value fit in from Fig. \ref{fig:collapse} for keratocytes (no additional fit is done in this figure), and $E = 600$ mV/mm. $d = 2$. {The broad distribution of directionalities seen in experiment, including points near $-1$, is expected, and also seen in our simulations (Appendix \ref{app:distributions_constant}). } }
    \label{fig:radius}
\end{figure}

\section*{Model suggests accuracy moderately depends on cell size}
In the experimentally-relevant range of fields, $\mathcal{I}(\psi) \sim \kappa^2 \sim R^2$ depends on radius. Is this size-dependence detectable? The best-available data on accuracy as a function of cell size is \cite{sun2013keratocyte} on keratocytes. However, in these experiments, cell areas vary only $\sim$twofold. In Fig. \ref{fig:radius} we compare our bound (Eq. \eqref{eq:periodic_bound}) with data from \cite{sun2013keratocyte}. We start with $\gamma$ fit for keratocytes from Fig. \ref{fig:collapse}. Since we cannot separate $\beta$ and $N$ in our fit, we pick $N = 10^4$, and set $\beta_0 = \gamma \sqrt{d/N} = \mu R_0 /D$, where $R_0$ is an average keratocyte radius for the experiments in Fig. \ref{fig:radius}. We plot directionality $\langle \cos (\hat{\psi}-\psi) \rangle$ as a function of increasing radius $R$ relative to $R_0$. This result is insensitive to $N$ (Appendix \ref{app:radius_moved}). We also plot experimental measurements \cite{sun2013keratocyte}, computing effective radius as $R = \sqrt{A/\pi}$ and defining $R_0$ as the average keratocyte radius. The experimental data is scattered, and does not show a clear increase with radius, even when averaged into two groups (above- and below-average size cells). However, the experimental results are also too noisy to rule out our predicted size-dependence.
 We have held $N$ constant while varying $R$ in Fig. \ref{fig:radius}; we see a slightly stronger dependence on $R$ if sensor density $N/(4\pi R^2)$ is held constant (Appendix \ref{app:scaling}).
 $\kappa \ll 1$ in Fig. \ref{fig:radius}, and so as in Fig. \ref{fig:circular_variance_simple},  $\langle \cos (\hat{\psi}-\psi)\rangle$ for the maximum likelihood estimator is smaller than Eq. \eqref{eq:periodic_bound}, but has a similar dependence on $R$ (Appendix \ref{app:radius_moved}). 
 
 {Additional measurements of directionality as a function of cell size would provide a rigorous check on our predictions. However, because of the relatively small range of sizes seen in the keratocyte experiments, the difference between directionality of small and large cells will be small. For the parameters in Fig. \ref{fig:radius}, the predicted difference between the directionality of a cell with radius 14 microns (the larger group average) and a directionality of a cell with radius 10.5 microns (the smaller group average) is at most 0.11 across the reasonable range of electric fields. Since measurements of keratocyte directionality typically have error bars of $\sim \pm 0.1$ for $\sim 50$ cells, we'd expect to need $>100$ cells per group to show that large cells are more directed than small cells. Distinguishing different scaling laws (e.g. in Appendix \ref{app:scaling}) would require even more data.}

 {Cell-size-dependence of responses has also been observed in a different context -- the time taken to respond to a changed signal \cite{allen2013electrophoresis}.} %

{\section*{Brownian dynamics simulation of sensor diffusion and electrophoresis}
To understand how the cell will respond to dynamic signals, or to understand how the cell's directionality is correlated over time, we will need to simulate how the sensor configuration evolves over time -- to generate stochastic trajectories of sensors on the surface of the cell. To do this, we will use the stochastic differential equation corresponding to Eq. \eqref{eq:continuity_general}. First, we rescale our units of time to $\tilde{t} = t/\tau_\textrm{forget}$, where $\tau_\textrm{forget} \equiv R^2/D$ is the time for proteins to spread over the cell by diffusion in the absence of a field, finding
\begin{equation}
    \frac{\partial}{\partial \tilde{t}} p(\theta,\tilde{t}) = -\frac{\partial}{\partial \theta}\left[-\kappa \sin (\theta-\psi)
    p(\theta,\tilde{t})\right] + \frac{\partial^2}{\partial \theta^2} p(\theta,\tilde{t}), \label{eq:rescaled_continuity}
\end{equation}
where $\kappa=\mu ER/D$. We will work in these rescaled units, always considering averaging times or exposure times relative to the forgetting time $\tau_\textrm{forget}$, which is the relevant timescale in the problem.
The stochastic differential equation (``Langevin equation'') corresponding to Eq. \eqref{eq:rescaled_continuity} is
\begin{equation}
    \dfrac{\mathrm{d}}{\mathrm{d}\tilde{t}} \theta_i(\tilde{t}) = - \kappa \sin[ \theta_i(\tilde{t})-\psi(\tilde{t})] + \xi(\tilde{t}), \label{eq:langevin}
\end{equation}
where $\xi(\tilde{t})$ is a Gaussian Langevin noise with $\langle \xi(\tilde{t}) \xi(\tilde{t}')\rangle = 2 \delta(\tilde{t}-\tilde{t}')$ and the field direction $\psi(\tilde{t})$ can change with time. The sensor positions will, once simulated for a long enough time in a constant field, have a steady-state probability distribution $p(\theta)$ given by Eq. \eqref{eq:Smoluchowski_ss}. We then simulate $N$ independent sensors diffusing in the presence of the electric field by integrating Eq. \eqref{eq:langevin} with the Euler-Maruyama method. Unless stated elsewhere, we use time step $\Delta \tilde{t}=0.001$.

\section*{Cells benefit from time-averaging over times $T>R^2/D$}
Cells or groups of cells sensing chemical concentrations or gradients may improve accuracy by integrating multiple measurements of the signal over a time 
$T$ \cite{berg1977physics,ten2016fundamental,hu2010physical,hu2011geometry,camley2017cell,camley2018collective,hopkins2020chemotaxis}, reducing variance by $N_\textrm{meas} \sim T/\tau_\textrm{corr}$, where $\tau_\textrm{corr}$ is the measurement correlation time. The relevant correlation time for galvanotaxis is $\tau_\textrm{forget} \equiv R^2/D$. To test the effect of integrating multiple measurements in galvanotaxis, we use Brownian dynamics simulations using Eq. \eqref{eq:langevin}. We then compute the time-averaged direction of the estimator of the field as $\boldsymbol\rho_T \equiv \frac{1}{T}\int_{t}^{t+T} dt' \boldsymbol\rho(t')$, where  $\bm{\rho}(t) = \sum_i \left(\cos \theta_i(t),\sin \theta_i(t)\right)$ is the sum of sensor positions. 
From the estimator of the field direction $\boldsymbol\rho_T(t)$, we compute the circular variance $V_T$ of the directions of the time-averaged estimator. Averaging does decrease error over a broad range of $\beta E$ (Fig. \ref{fig:timeaverage}a), though for $\beta E \lesssim 0.05$ this may be masked, since even with averaging, $V_T$ is near its maximum of 2. As expected, for $T \ll \tau_\textrm{forget}$, the circular variance $V_T$ limits back to the case of zero averaging time (Fig. \ref{fig:timeaverage}b). For $T \gg \tau_\textrm{forget}$, $V_T \sim \tau_\textrm{forget}/T$, as we would expect if $V_T \approx V_{T=0} / N_\textrm{meas}$ with the number of independent measurements $N_\textrm{meas} \sim T/\tau_\textrm{forget}$.

\begin{figure}
    \centering
    \includegraphics[width=\columnwidth]{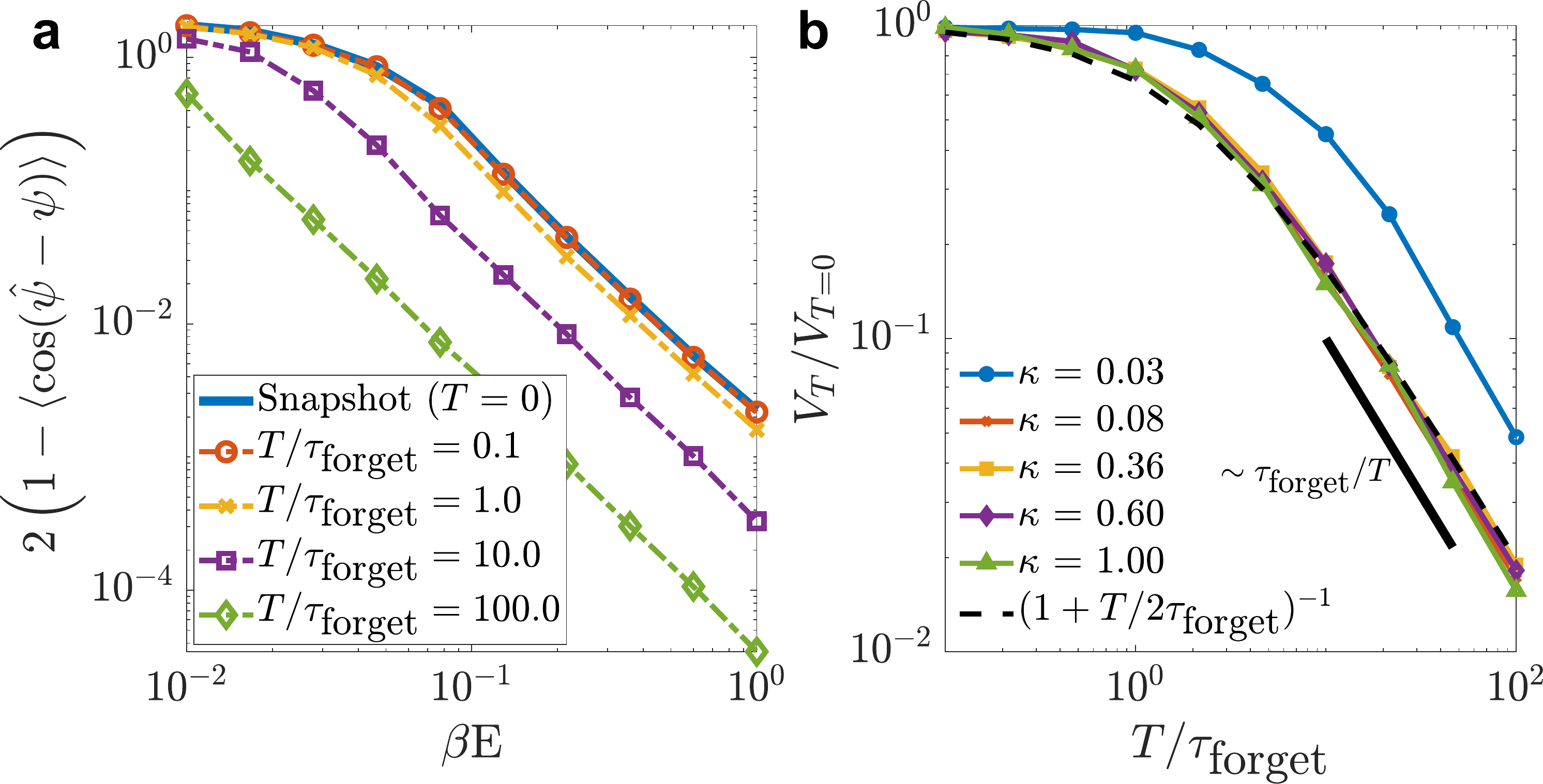}
    \caption{To reduce error, cells must average over times long compared with $\tau_\textrm{forget}$. {\bf a:} Circular variance $V_T$ as a function of $\kappa = \beta E$ for several averaging times $T$. {\bf b:} $V_T$ rescaled by its snapshot value $V_{T=0}$ behaves as $\sim \tau_\textrm{forget}/T$ at large $T$. Lines are an average of 10 simulations of length $1000 \tau_\textrm{forget}$,  with $\Delta t = 0.01 \tau_\textrm{forget}$. We can capture most time-averaging effects by a scaling form 
    $\dfrac{V_T}{V_{T=0}} = \left(1 + \frac{T}{2\tau_\textrm{forget}}\right)^{-1}$,  which we chose to match the expected asymptotic forms. Here $V_{T=0}$ is the circular variance with no time averaging. This collapse fails at small $\kappa$ because $V_T$ reaches its maximum value of $2$. Simulation length was set to $(0.1T/\tau_\textrm{forget})\times1000\tau_\textrm{forget}$ for $T/\tau_\textrm{forget}>10$, capping at $5000\tau_\textrm{forget}$.}
    \label{fig:timeaverage}
\end{figure}

In strong contrast to chemotaxis and concentration sensing, it may be difficult for galvanotaxing cells to gain accuracy by time-averaging over $T \gtrsim \tau_\textrm{forget}$. Keratocytes have $\tau_\textrm{forget} \sim 15$ min \cite{allen2013electrophoresis}, making time-averaging unlikely as keratocytes respond to field changes within a few minutes \cite{allen2013electrophoresis}. By comparison, estimates of averaging time for Dictyostelium chemotaxis range from $2-20$ s \cite{fuller2010external,segota2013high,van2007biased}. However, {\it in vivo}, fibroblasts can take hours to respond to electric fields from wounds \cite{guo2010effects}, making large $T$ plausible, while smaller-radius cell types have shorter $\tau_{\textrm{forget}} = R^2/D$. The utility of time-averaging is context- and cell-type-dependent. Our results show that the constraints on useful time averaging are qualitatively different for galvanotaxing cells than for chemotaxing cells. %

\begin{figure*}[ht]
    \centering
    \includegraphics[width=\textwidth]{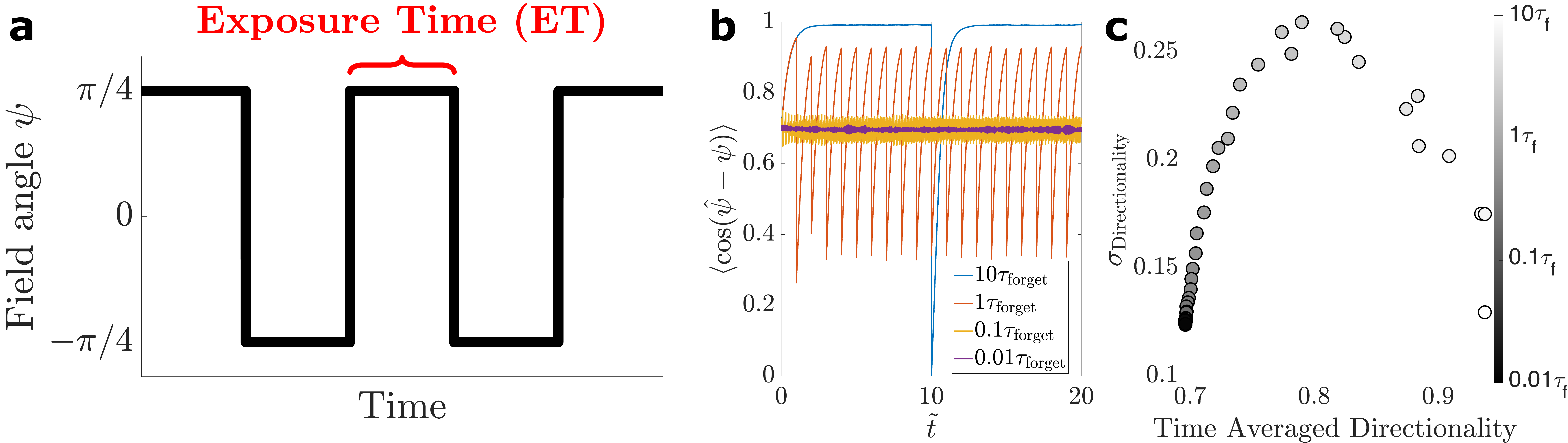}
    \caption{ Response to switching fields presents tradeoff between accuracy and variability. $\mathbf{a}$: Schematic of switching fields over time. $\mathbf{b}$: Directionality as function of dimensionless time $\tilde{t}=t/\tau_\mathrm{forget}$ for different exposure times ET. $\mathbf{c}$: Standard deviation of instantaneous directionality $\sigma_\mathrm{Directionality}$ (per cell) and time averaged directionality (average of $\langle\cos(\hat\psi-\psi)\rangle$ from $\tilde{t}=10$ to $\tilde{t}=20$) vary as ET is varied (color bar). Averages are over 1000 cells, $N = 1000$ sensors, and $\kappa=0.358$. An equilibration time of $4\tau_\mathrm{forget}$ in a constant field ($\psi=0$) was simulated prior to collecting data.} 
    \label{fig:switching}
\end{figure*}

\section*{Switching fields present tradeoff between accuracy and variance}
 {So far, we have studied cells in constant electric fields, but controlling precisely where a cell goes may require more complex, changing fields. We are motivated in particular by the work of Zajdel et al., who  developed an experimental setup for galvanotaxis with two pairs of electrodes, allowing for application of electric fields in two perpendicular directions \cite{zajdel2020scheepdog}. By rapidly switching the electric field between the $+x$ and $+y$ direction every 10 seconds, Zajdel et al. found that cells could be guided along the 45-degree diagonal \cite{zajdel2020scheepdog}. How does the large time for sensors to diffuse across the cell $\tau_\textrm{forget}$ influence responses to rapidly-switched fields? Using our Brownian dynamics simulations, we switch field direction $\psi(t)$ over two orthogonal directions, $\pm\pi/4$, varying the ``exposure time'' ET over which the field is constant (Fig. \ref{fig:switching}a). For $\textrm{ET}\ll \tau_\textrm{forget}$, sensors cannot rearrange on the cell surface as quickly as the field switches, leading the cell to compromise between the two directions $\psi = \pm \pi/4$, and travel in the average field direction $\psi=0$ -- precisely as found in experiments controlling groups of keratinocytes (see Fig. 4d of \cite{zajdel2020scheepdog}).} 
 
{ We measure how precisely the cell is following the current field $\mathbf{E}(t)$ by the instantaneous directionality, the cosine of the angle between the cell's direction and the current field (Fig. \ref{fig:switching}b). For rapid switching ($\textrm{ET}\ll \tau_\textrm{forget}$) the cell has low accuracy to the instantaneous field direction -- but also relatively low variability, i.e. the cell is going consistently in one direction, but not the instantaneous field direction $\psi(t)$ (Fig. \ref{fig:switching}b). However, at intermediate $\textrm{ET} \sim \tau_\textrm{forget}$, larger oscillations in directionality $\langle \cos (\hat{\psi}-\psi(t))\rangle$ appear as the sensors have time to repolarize in response to the changed field (Fig. \ref{fig:switching}b). In the regime $\textrm{ET} \lesssim \tau_\textrm{forget}$,  there is a clear tradeoff in control: increasing ET increases both directionality and variability in directionality (Fig. \ref{fig:switching}c). This means that, though -- on average -- the cell is more likely to be going in the desired direction $\psi(t)$, the large variation means that it is also more likely to in a different direction altogether. This is not a global tradeoff, though. For $\textrm{ET}\gg\tau_\textrm{forget}$, variability of directionality decreases (Fig. \ref{fig:switching}c). In this case, the distribution of $\cos(\hat\psi-\psi({t}))$ is heavily skewed (Appendix \ref{app:switching}). We also find similar results, though with larger fluctuations, when we switch fields between $0$ and $\pi$ (Appendix \ref{app:switching}).}
 
{The tradeoff between average directionality and variability of directionality (Fig. \ref{fig:switching}c) and the appearance of large oscillations in the directionality (Fig. \ref{fig:switching}b), are new predictions that arise as a signature of the long time for sensors to redistribute across the membrane in the electric field.}  %

\section*{Galvanotaxis and chemotaxis may share similar sensing strategy}
Our results show that circular or spherical cells can measure the direction of an electrical field by summing the vectors pointing to their electromobile sensors $\boldsymbol{\rho}$. This direction can be found by the cell by local protrusions in the normal direction; see Appendix \ref{app:protrusionmodel}. This method of choosing a direction is exactly analogous to the estimator for chemotaxis of circular cells in \cite{hu2010physical,hu2011geometry}, where cells move toward the vector sum of {\it bound} sensors. This is unexpected, given essential differences between chemotactic and galvanotactic models: galvanotactic sensors reorganize and chemotactic ones do not, while galvanotactic sensors do not bind external ligand. Highly accurate processing of galvanotactic information and chemotactic information could then be performed by signaling networks shared between galvanotactic and chemotactic responses. Supporting this idea, chemotactic and galvanotactic response in Dictyostelium share  common core elements, including TORC2 and PI3K  \cite{gao2015large}. 

\section*{Further experimental tests}
A key implication of our modeling is that observed galvanotactic accuracies are physically plausible either with relatively few sensors ($\sim 100$) as  responsive as tdTomato-GPI, or with a larger number of sensors whose redistribution need not be obvious even in strong ($1000 $ mV/mm) electric fields. Our work provides a natural quantitative route to test a putative sensor by modifying its abundance (via knockdown or overexpression) or polarization (via changes in extracellular viscosity or pH \cite{kobylkevich2018reversing,sarkar2019electromigration}, or molecular charge), and then measure the cell's directionality as a function of field. Our work provides a quantitative prediction for how cell directionality depends on sensor abundance and polarization. We can also predict directionality as a function of cell size (Fig. \ref{fig:radius}) and directionality as a function of exposure time in a switching field (Fig. \ref{fig:switching}) for any cell type once $\gamma$ is fit from the universal directionality-field strength curve (Fig. \ref{fig:collapse}).

\section*{Sensor interactions and cell shape caveats}
We have neglected sensor-sensor interactions by assuming that sensor positions are independent from one another. We believe sensor-sensor interactions are likely to be relevant only at very large fields $E$ and very large $N$. If sensors were uniformly spread over the spherical cap within $\pi/4$ of the field  (this is more concentrated than $\kappa = 1$, corresponding to electric fields much larger than experimentally reasonable), sensor density would be $N/A$ with $A = 2 \pi R^2(1-\cos{(\pi/4)}) \approx 1.8 R^2$ the area of the cap. With $N = 10^4$ sensors and $R = 5~\mu\textrm{m}$, typical distance between sensors is $\sqrt{A/N} \approx 70$ nm, beyond typical screening lengths. Hydrodynamic interactions in a membrane may be long-range \cite{noruzifar2014calculating,oppenheimer2009correlated,oppenheimer2011plane}, experimentally measured at micron scale \cite{chein2019flow}. Including hydrodynamic interactions would not alter the steady-state $p(\theta)$, but would lead to  correlated sensor diffusion, changing the time required for a sensor configuration to decorrelate, likely only altering the time scale $\tau_{\textrm{forget}}$.

 Changing cell shape would also affect our results, especially since the electromigration velocity depends on how the local field is oriented with respect to the membrane surface. This will allow elongated cells to have different sensitivities to fields parallel and perpendicular to them, as in chemotaxis \cite{hu2011geometry}. {Initial calculations with an elliptical geometry, which will be published in a separate manuscript, show that an ellipse with semi-minor and semi-major axes $R_1$ and $R_2$ with an aspect ratio between 2 and 3 will have a Fisher information within 16--24\% of a circular cell with radius $(R_1+R_2)/2$. Using the full formula for an elliptical cell would not change the curve in Fig. \ref{fig:collapse}, but would change the fitted $\beta$ values by 8--11\%. We thus expect our approach to be quite acceptable even for fairly elongated cells.} %

\begin{acknowledgments}
We thank Wei Wang, Amit R. Singh, and Wouter-Jan Rappel for useful feedback and discussions. This material is based upon work supported by the National Science Foundation under Grant No. MCB 2119948 and PHY 1915491.
\end{acknowledgments}

\renewcommand{\thefigure}{S\arabic{figure}}
\setcounter{figure}{0}

\onecolumngrid
\appendix

\section*{Supplementary Information: ``Physical limits on galvanotaxis''}
\begin{figure}[htb]
    \centering
    \includegraphics[width=0.3\textwidth]{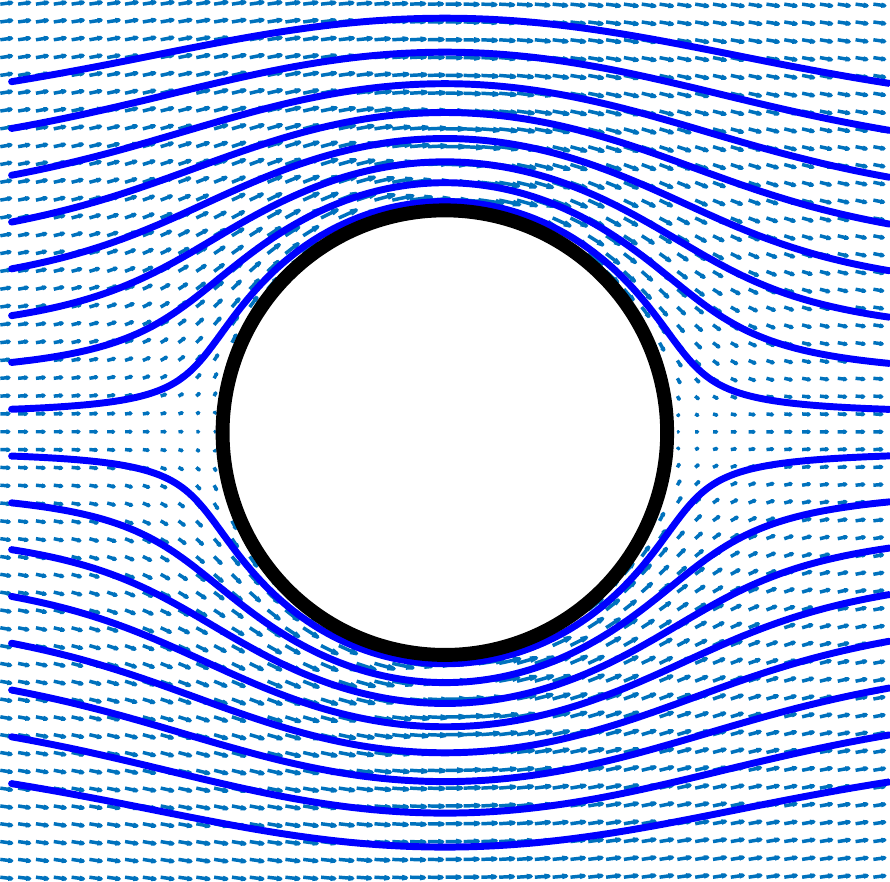}
    \caption{{Field lines and vector fields for electric field around the boundary of the cell if we assume $\partial V/\partial r = 0$ at the membrane, as in \cite{mclaughlin1981role}, though this assumption is not necessary (see text).}}
    \label{fig:fieldlines}
\end{figure}

{\section{\label{app:efield_details}Details of electric field assumptions}
In the main text, we have written our assumption that the velocity of the sensor is
\begin{equation}
\mathbf{v} = \mu \left[\mathbf{E} -  (\mathbf{E}\cdot \hat{\mathbf{n}}) \hat{\mathbf{n}}\right],
\end{equation}
where $\mathbf{E}$ is the externally-applied field -- i.e. the field far away from the cell -- and $\mathbf{n}$ the normal to the cell surface. Essentially, this says that the motion of the sensor is only in the tangential plane of the membrane (we subtract off the normal component) -- which will be true as long as the membrane is not deforming. $\mu$ is an effective mobility akin to that derived by \cite{mclaughlin1981role}. This parameter $\mu$ could depend on the zeta potential of the surface and the extracellular piece of the sensor, the viscosity of the fluid near the membrane and the membrane viscosity as well as the sensor geometry. The assumptions of \cite{mclaughlin1981role} incorporate both electrophoresis and electroosmotic flow. However, our focus is not on the details of these specific assumptions -- we view $\mu$ as a phenomenological parameter, essentially describing the linear response to an electric field, which we set from experiment. However, we do want to mention one issue: why have we written this in terms of the {\it applied} electric field $\mathbf{E}$? The electric field near the cell is not the same as the applied electric field $\mathbf{E}$ (Fig. \ref{fig:fieldlines}), but the tangential component of the field at the surface of the cell is {\it proportional} to the tangential component of the external field.}

{For a simple example, let's think about modeling a circular cell in an applied electric field $E_0 \mathbf{\hat{x}}$. The potential $V(r,\theta)$ will obey Laplace's equation $\nabla^2 V = 0$ external to the cell, and will have the potential far from the cell go to $V \to -E_0 x = -E_0r\cos\theta$. For simplicity, we start with a boundary condition $\partial V/\partial r|_{r=R} = 0$ (this is used by \cite{mclaughlin1981role}, and corresponds to an assumption that the cell membrane has negligible conductivity in comparison to the cell interior and the fluid outside the cell \cite{kotnik2000analytical,pucihar2009time}). By using the general solution to Laplace's equation in two dimensions \cite{griffiths2005introduction}, we can find $V(r,\theta) = -E_0 \left(\frac{R^2}{r}+r\right)\cos\theta$. The electric field as a function of position $\mathbf{E}(\mathbf{r})$ is then 
\begin{align}
   \mathbf{E}=-\nabla V=-\dfrac{\partial V}{\partial r}\mathbf{\hat{r}}-\dfrac{1}{r}\dfrac{\partial V}{\partial\theta}\bm{\hat{\theta}}&=E_0\left[\left(-\dfrac{R^2}{r^2}+1\right)\cos\theta\mathbf{\hat{r}}+\left(\dfrac{R^2}{r^2}+1\right)(-\sin\theta)\bm{\hat{\theta}}\right].
\end{align}
We plot this in Fig. \ref{fig:fieldlines}. The tangential component (the $\bm{\hat{\theta}}$ component) at the cell boundary is $-2E_0\sin\theta$. The tangential part of the electric field {\it at the cell membrane} is thus proportional to the tangential part of the external field, suggesting our assumption $(\mathbf{v}=-\mu E\sin\theta\bm{\hat\theta})$ with $E$ the external electric field is reasonable. The boundary condition $\partial V/\partial r|_{r=R} = 0$ is the one used by \cite{mclaughlin1981role}, but many different boundary conditions, e.g. a finite conductivity, or treating the cell and its environment as uniform dielectric materials, will still lead to proportionality of this sort -- but with a different prefactor (see, e.g. \cite{cole1972membranes,griffiths2005introduction} for the sphere case).} 

\section{\label{app:sphere} Spherical cell geometry}

The concentration of sensors on a spherical cell with an electric field in the $z$ direction is worked out in \cite{mclaughlin1981role} as $c \sim e^{\beta E \cos \theta}$. We can generalize this to a field in an arbitrary direction as $p(\theta,\phi) \sim \exp\left(\beta \mathbf{E}\cdot \hat{\mathbf{u}}\right)$, where $\hat{\mathbf{u}} = \left(\cos\phi \sin\theta,\sin\phi\sin\theta,\cos\theta\right)$ is the unit vector on the sphere {and the field $\mathbf{E}=E(\cos\psi_\phi \sin\psi_\theta,\sin\psi_\phi\sin\psi_\theta,\cos\psi_\theta)$} -- this is the von Mises-Fisher distribution \cite{mardia2000directional}. To model a cell on a substrate, where velocities can only be measured in the plane of the substrate, we will assume explicitly that the field is in the $xy$ plane {$(\psi_\theta=\pi/2)$}. {In a 3D system, e.g. a single cell in extracellular matrix or in solution, the cell would have to estimate both the azimuthal and polar angle of the electric field. However, wound healing and galvanotaxis are often on flat substrates, where the cell has other information about the location of the substrate (e.g. apicobasal polarity in epithelia). Thus, we assume that  the polar angle is fixed and the cell only needs to estimate the azimuthal angle $\psi$. This reduces the estimation on a sphere to a single $2\pi$-periodic variable, so our modified Cramer-Rao bound should hold.} We can then, considering the {azimuthal field angle} as $\psi$, write the probability distribution of a single sensor as $p(\phi,\theta) = Z^{-1} \exp{[\kappa \cos (\phi-\psi)\sin\theta]}$. For $\int \mathrm{d}\theta\sin\theta d\phi p(\phi,\theta) = 1$, we have $Z = (4\pi\sinh{\kappa})/\kappa$.
Then, the log-likelihood takes the form
\begin{equation}
\label{eq:sphereloglike}
    \ln \mathcal{L}(\psi,E;\bm{\theta}) =
    N \ln \left[\dfrac{\kappa}{4\pi\sinh{\kappa}}\right] + \kappa \sum_{i=1}^N \cos (\phi_i - \psi)\sin\theta_i.
\end{equation}
We note that the second term on the right can be rewritten, so that the log-likelihood is 
\begin{align}
    \ln \mathcal{L}(\psi,E;\bm{\theta}) &=
    N \ln \left[\dfrac{\kappa}{4\pi\sinh{\kappa}}\right] + \beta \sum_{i=1}^N \mathbf{E} \cdot \hat{\mathbf{u}}^{xy}_i \\
    &=
    N \ln \left[\dfrac{\kappa}{4\pi\sinh{\kappa}}\right] + \beta \mathbf{E} \cdot \boldsymbol\rho^{xy},
\end{align}
where $\hat{\mathbf{u}}^{xy}_i = (\cos \phi_i \sin\theta_i,\sin\phi_i\sin\theta_i,0)$ and $\boldsymbol\rho^{xy} = \sum_i \hat{\mathbf{u}}^{xy}_i$ is the sum of the sensor locations -- projected into the xy plane. 
The orientation of the field $\mathbf{E} = E(\cos \psi,\sin\psi,0)$ that maximizes the likelihood is the one that puts $\mathbf{E}$ in line with $\boldsymbol\rho^{xy}$, exactly as in the 2D circle case; (we can also see this by explicitly differentiating the log-likelihood; see Appendix \ref{app:MLE}.)

\begin{figure}[htb]
    \centering
    \includegraphics[width=0.6\columnwidth]{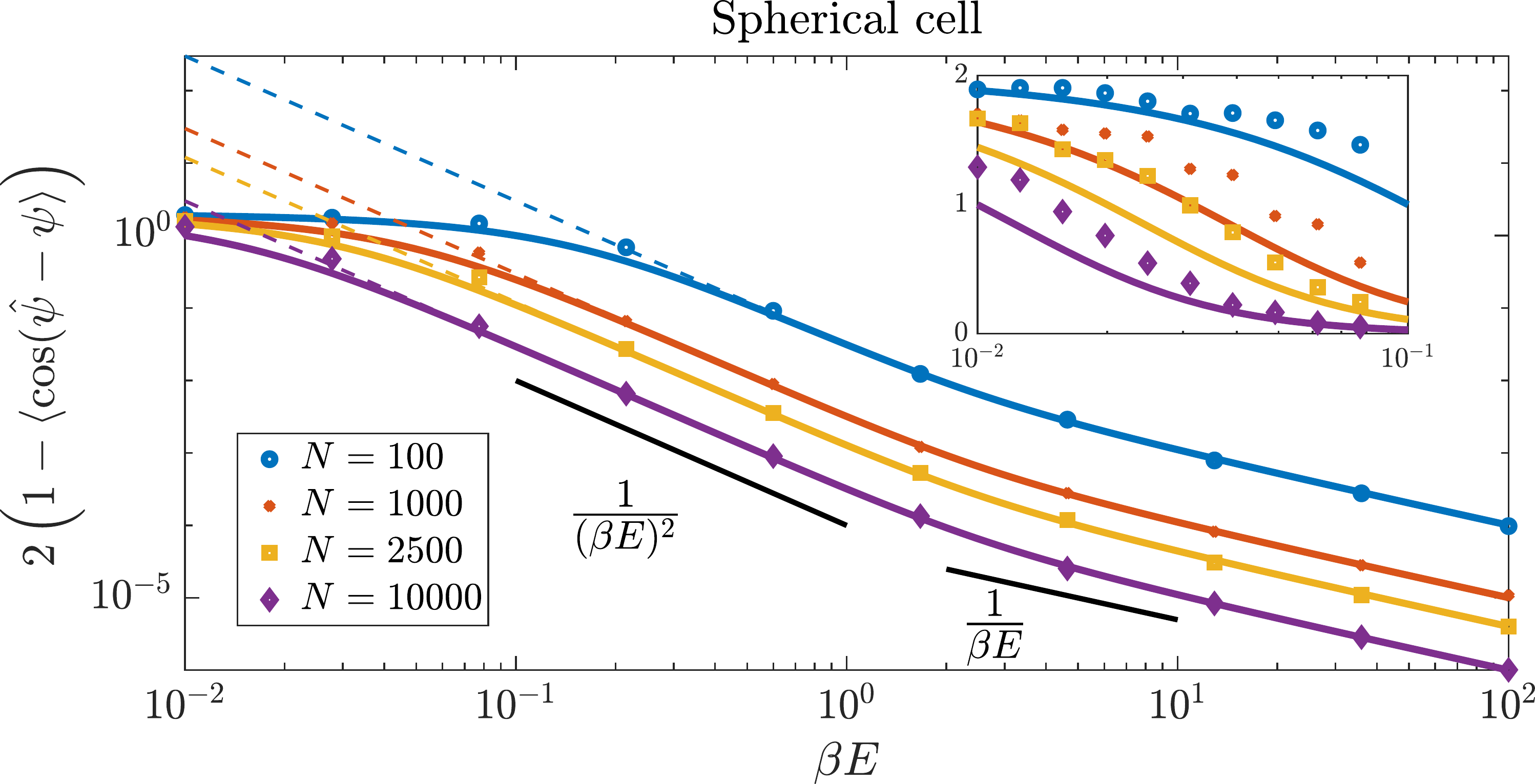}
    \caption{Accuracy of galvanotactic estimation as a function of electric field and number of sensors for a spherical cell. Solid lines are the simple periodic bound (Eq. (8) in the main text), dashed lines are the normal Cramer-Rao bound (Eq. (7) in the main text), symbols are computed from stochastic simulation. Inset shows zoomed in region in linear spacing, showing MLE variance is systematically above the periodic bound.}
    \label{fig:spherical_variance_simple}
\end{figure}

Now we can compute the Fisher information by first calculating the second derivatives:
\begin{align}
    \left\langle\frac{\partial^2}{\partial \psi^2} \ln \mathcal{L}(\psi,E;\bm{\theta})\right\rangle &= -\kappa \sum_{i=1}^N \langle \cos (\phi_i - \psi)\sin{\theta_i} \rangle, \\
    \left\langle\frac{\partial^2}{\partial \psi \partial E} \ln \mathcal{L}(\psi,E;\bm{\theta})\right\rangle &= \beta \sum_{i=1}^N \langle \sin (\phi_i - \psi)\sin{\theta_i} \rangle, \\
    \left\langle\frac{\partial^2}{\partial E^2} \ln \mathcal{L}(\psi,E;\bm{\theta})\right\rangle &= -N\beta^2\dfrac{\sinh^2\kappa-\kappa^2}{\kappa^2\sinh^2{\kappa}}.
\end{align}
The expectation values $\langle\cos{(\phi_i-\psi)}\sin\theta_i\rangle=\kappa^{-1}(\kappa\coth\kappa-1)$ and $\langle\sin{(\phi_i-\psi)}\sin\theta_i\rangle=0$ by symmetry.
By taking the negatives of the calculated derivatives, this gives us the final Fisher information matrix $\bm{\mathcal{I}}$,
\begin{equation}
    \bm{\mathcal{I}} = 
    N\begin{pmatrix}
 \kappa \coth\kappa-1 & 0 \\[5pt]
0 & \beta^2\dfrac{\sinh^2\kappa-\kappa^2}{\kappa^2\sinh^2{\kappa}}
\end{pmatrix}.
\end{equation}
The upper-left component of the matrix corresponds to $\mathcal{I}(\psi)$, which was given as Eq. (6) in the main text. 

We show the spherical-cell errors as a function of $\beta E$ and $N$ in Fig. \ref{fig:spherical_variance_simple}. Here, the stochastic simulation is done by generating random sensors according to $p(\phi,\theta) \sim Z^{-1} \exp\left[\kappa \cos(\phi-\psi)\sin\theta\right]$ by rejection sampling (we note that to sample from this distribution on the sphere, we are sampling from $p(\phi,\theta) \sin \theta \mathrm{d}\theta \mathrm{d}\phi$). For each parameter, we generate 1000 sensor configurations. For each configuration, we compute the estimator $\hat{\psi}$ by summing sensor positions to find $\boldsymbol\rho^{xy}$, and then use this to compute $V$.

\section{\label{app:MLE} Computing maximum likelihood estimators for field direction and strength and corresponding Fisher informations}

For the circular cell model, the probability density of a single sensor is $p(\theta) = Z^{-1} e^{\kappa \cos (\theta-\psi)}$. For $p(\theta)$ to integrate to one, $Z = 2 \pi I_0(\kappa)$. We note $\kappa = \beta E$. Then the log-likelihood is,
\begin{equation}
\label{eq:circleloglike}
    \ln \mathcal{L}(\psi,E;\bm{\theta}) = - N \ln \left[2\pi I_0(\kappa)\right] + \kappa \sum_{i=1}^N \cos (\theta_i - \psi).
\end{equation}

We discuss in the main text that {the maximum-likelihood estimation of the electricial field direction is to sum the vectors pointing in the direction of their electromobile sensors.} Here we show this a little more explicitly, as well as showing how estimators for the electric field magnitude can be derived for the circle and the sphere. 
The log-likelihood functions for a circular and spherical cell are given by Eqs. \ref{eq:circleloglike} and \ref{eq:sphereloglike}, respectively. To find the maximum likelihood estimator, we find the value $\hat{\psi}$ that maximizes these log-likelihoods by differentiating each equation with $\psi$ and setting this equal to zero,
\begin{align}
    \mathrm{Circle:}~\left.\dfrac{\partial}{\partial \psi} \ln \mathcal{L}(\psi,E;\bm{\theta})\right|_{\psi = \hat{\psi}} &= \kappa \sum_{i=1}^N \sin (\theta_i - \hat{\psi}) = 0,\\
    \mathrm{Sphere:}~\left.\dfrac{\partial}{\partial \psi} \ln \mathcal{L}(\psi,E;\bm{\theta})\right|_{\psi = \hat{\psi}} &= \kappa \sum_{i=1}^N \sin (\phi_i - \hat{\psi})\sin \theta_i = 0.
\end{align}
For the circle, we can solve for $\hat{\psi}$ by using the trigonometric identity $\sum\sin{(\theta_i-\hat{\psi})}=\sum(\sin \theta_i\cos \hat{\psi}-\cos \theta_i\sin \hat{\psi})=0$. Separating the two terms allows us to factor out $\hat{\psi}$ from the summation: $\cos \hat{\psi}\sum\sin \theta_i=\sin \hat{\psi}\sum\cos \theta_i$. From this juncture, we can then solve for the estimator, $\hat{\psi}$, of the field direction. An analogous calculation can be done for the sphere. These results are:
\begin{align}
    \mathrm{Circle:}~\tan{\hat{\psi}} &= \dfrac{\sum_{i=1}^N\sin\theta_i}{\sum_{i=1}^N\cos\theta_i}, \\[10pt]
    \mathrm{Sphere:}~\tan{\hat{\psi}} &= \dfrac{\sum_{i=1}^N\sin{\phi_i}\sin\theta_i}{\sum_{i=1}^N\cos{\phi_i}\sin\theta_i}.
\end{align}
We see that the angle of the maximum likelihood estimator is obtained from the components of the summed vector of the sensor locations, $\boldsymbol\rho$.

In addition, we can find the maximum likelihood estimator for the field strength $E$.
Differentiating the log-likelihood functions for the circle and sphere for the field strength, $E$, yields:
\begin{align}
    \mathrm{Circle:}~\left.\dfrac{\partial}{\partial E} \ln \mathcal{L}(\psi,E;\bm{\theta})\right|_{E = \hat{E}} &= \beta \sum_{i=1}^N \cos (\theta_i - \psi) - \beta  N\dfrac{I_1(\hat{\kappa})}{I_0(\hat{\kappa})} = 0,\\
    \mathrm{Sphere:}~\left.\dfrac{\partial}{\partial E} \ln \mathcal{L}(\psi,E;\bm{\theta})\right|_{E = \hat{E}} &= \beta \sum_{i=1}^N \cos (\phi_i - \psi)\sin \theta_i+\dfrac{N}{\hat{E}}-\dfrac{N\beta}{\tanh{\hat{\kappa}}} = 0.
\end{align}
where $\hat{\kappa} = \beta \hat{E}$. We can then reorganize these equations to provide formulas for the maximum likelihood estimators. 
\begin{align}
    \mathrm{Circle:}~\dfrac{I_1(\hat\kappa)}{I_0(\hat\kappa)} &= \dfrac{1}{N}\sum_{i=1}^N \cos (\theta_i - \psi),\\
    \mathrm{Sphere:}~\dfrac{1}{\hat\kappa}(\hat\kappa\coth{\hat\kappa}-1) &= \dfrac{1}{N}\sum_{i=1}^N \cos (\phi_i - \psi)\sin \theta_i.
\end{align}
We cannot analytically solve for the electric field here, but this provides a straightforward numerical way to find the maximum likelihood estimator $\hat{E}$.

In the main paper, we have introduced the Fisher information for the field direction only. However, it is straightforward to extend the results to describe simultaneous estimation of the field direction and strength. Simultaneously estimating the field strength does not change the accuracy of the directional estimate -- as we might guess, because the estimator for the field direction -- the sum of sensor directions -- doesn't depend on field strength. 

 Then we can work out the Fisher information matrix components {$\mathcal{I}_{\alpha\beta} = -\left\langle \frac{\partial^2 \ln \mathcal{L}}{\partial \alpha \partial \beta}\right\rangle $} where $\alpha,\beta$ are dummy variables indicating field magnitude $E$ or angle $\psi$. To do this, we compute:
\begin{align}
    \left\langle\frac{\partial^2}{\partial \psi^2} \ln \mathcal{L}(\psi,E;\bm{\theta})\right\rangle &= -\kappa \sum_{i=1}^N \langle \cos (\theta_i - \psi) \rangle, \\
    \left\langle\frac{\partial^2}{\partial \psi \partial E} \ln \mathcal{L}(\psi,E;\bm{\theta})\right\rangle &= \beta \sum_{i=1}^N \langle \sin (\theta_i - \psi) \rangle, \\
    \left\langle\frac{\partial^2}{\partial E^2} \ln \mathcal{L}(\psi,E;\bm{\theta})\right\rangle &= -N\frac{\beta^2}{2}\left[1 + \frac{I_2(\kappa)}{I_0(\kappa)} - 2 \left(\frac{I_1(\kappa)}{I_0(\kappa)}\right)^2\right].
\end{align}
The expectation of the cosine is $\langle \cos(\theta_i - \psi) \rangle = I_1(\kappa)/I_0(\kappa)$ and $\langle \sin(\theta_i-\psi) \rangle = 0$ by symmetry.

This gives us the final Fisher information matrix $\bm{\mathcal{I}}$,
\begin{equation}
\bm{\mathcal{I}} = 
    N\begin{pmatrix}
 \kappa \dfrac{I_1(\kappa)}{I_0(\kappa)} & 0 \\
0 & \dfrac{\beta^2}{2}\left[1 + \dfrac{I_2(\kappa)}{I_0(\kappa)} - 2 \left(\dfrac{I_1(\kappa)}{I_0(\kappa)}\right)^2\right]
\end{pmatrix}.
\end{equation}
These results are known as the Fisher information associated with multiple samples of a von Mises distribution \cite{mardia2000directional}.

\section{\label{app:periodic_cramer}Periodic generalization of the Cramer-Rao bound}

We show here a brief derivation of the periodic Cramer-Rao bound of Eq. (8) in the main text. This result can also be directly derived from a variant of the periodic Cramer-Rao bound found in \cite{mardia2000directional}, Chapter 5, but we show a derivation here because it is relatively straightforward but not well known in the physics literature. We start with a standard derivation of the ordinary Cramer-Rao bound, similar to, e.g., \cite{kay1993fundamentals}. We use this here to show how the derivation of the periodic Cramer-Rao bound follows from similar logic.

\subsection{Ordinary Cramer-Rao bound derivation}

If we have a parameter $\psi$ that we want to estimate from data $\mathbf{x}$, we can construct an estimator $\hat{\psi}(\mathbf{x})$. If the data is generated from a process with a probability density for the data given the parameter $p(\mathbf{x} | \psi)$, then requiring that the estimator is unbiased is requiring $\langle \hat{\psi} \rangle = \psi$ for all $\psi$ i.e.
\begin{equation}
    \int \mathrm{d}\mathbf{x} (\hat{\psi}(\mathbf{x}) - \psi) p(\mathbf{x}|\psi) = 0. \label{eq:unbiased_regular}
\end{equation}
Because Eq. \eqref{eq:unbiased_regular} is true for all values of the parameter $\psi$, we can take a derivative of Eq. \eqref{eq:unbiased_regular}, finding:
\begin{equation}
    \int \mathrm{d}\mathbf{x} (\hat{\psi}(\mathbf{x}) - \psi) \frac{\partial}{\partial \psi} p(\mathbf{x}|\psi) - \int \mathrm{d}\mathbf{x} p(\mathbf{x}|\psi) = 0.
\end{equation}
The second term is just 1, since the probability density is normalized, so
\begin{align}
    1 &= \int \mathrm{d}\mathbf{x} (\hat{\psi}(\mathbf{x}) - \psi) \frac{\partial}{\partial \psi} p(\mathbf{x}|\psi), \\
     1 &= \int \mathrm{d}\mathbf{x} (\hat{\psi}(\mathbf{x}) - \psi) p(\mathbf{x}|\psi)\frac{\partial}{\partial \psi} \ln p(\mathbf{x}|\psi),\\
     1^2 &= \left[\int \mathrm{d}\mathbf{x} (\hat{\psi}(\mathbf{x}) - \psi) p(\mathbf{x}|\psi)\frac{\partial}{\partial \psi} \ln p(\mathbf{x}|\psi)\right]^2, \\
     1 &= \left[\int \mathrm{d}\mathbf{x} \left\{\sqrt{p(\mathbf{x}|\psi)} (\hat{\psi}(\mathbf{x}) - \psi)\right\}\left\{ \sqrt{p(\mathbf{x}|\psi)}\frac{\partial}{\partial \psi} \ln p(\mathbf{x}|\psi)\right\}\right]^2.
\end{align}
In the last line we have split $p$ into $\sqrt{p}\sqrt{p}$. Then we can apply the Cauchy-Schwarz inequality, $\left(\int \mathrm{d}\mathbf{x} f(\mathbf{x})g(\mathbf{x}) \right)^2 \le \left(\int \mathrm{d}\mathbf{x} f(\mathbf{x})^2 \right)\left(\int \mathrm{d}\mathbf{x} g(\mathbf{x})^2 \right)$, which gives
\begin{align}
    1 &\le \left[\int \mathrm{d}\mathbf{x} p(\mathbf{x}|\psi) (\hat{\psi}(\mathbf{x}) - \psi)^2 \right] \left[\int \mathrm{d}\mathbf{x} p(\mathbf{x}|\psi) \left(\frac{\partial}{\partial \psi} \ln p(\mathbf{x}|\psi)\right)^2\right].
\end{align}
Rearranging, this is
\begin{equation}
    \langle (\hat{\psi}-\psi)^2 \rangle \ge \frac{1}{\left\langle \left(\dfrac{\partial}{\partial \psi} \ln p(\mathbf{x}|\psi)\right)^2 \right\rangle}.
\end{equation}
The value $\left\langle \left(\frac{\partial}{\partial \psi} \ln p(\mathbf{x}|\psi)\right)^2 \right\rangle$ is $\mathcal{I}(\psi)$, the Fisher information for $\psi$, and can be equivalently written as $\left\langle \left(\frac{\partial}{\partial \psi} \ln \mathcal{L}\right)^2 \right\rangle = -\left\langle \frac{\partial^2}{\partial \psi^2} \ln \mathcal{L} \right\rangle$ where $\mathcal{L}(\psi ; \mathbf{x}) = p(\mathbf{x} | \psi)$ is the likelihood \cite{kay1993fundamentals}. This establishes the ordinary Cramer-Rao bound (Eq. (7) in the main text). 

\subsection{Generalization to periodic variables}

The ordinary Cramer-Rao bound fails for periodic variables in part because the normal definition of an unbiased estimator, $\langle \hat{\psi} \rangle = \psi$, will not be reasonable when the variable to be estimated is only defined modulo $2\pi$: $2\pi + \psi$ is just as good an estimate for $\psi$ as $\psi$ itself. We can define a generalized sense of ``unbiased," for a periodic variable, defining unbiased estimators $\hat{\psi}$ as having $\langle \sin (\hat{\psi}-\psi)\rangle = 0$ \cite{mardia2000directional}. While we note that any estimator we can think of as having an unbiased direction will have $\langle \sin (\hat{\psi}-\psi) \rangle$, it would be possible to construct an estimator where $\hat{\psi} = \psi + \pi$ that also satisfies this constraint. Other periodic generalizations of the Cramer-Rao bound exist, with different definitions of periodic unbiasedness, e.g. \cite{routtenberg2011periodic}. However, these generally require knowledge of the distribution of the estimator $\hat{\psi}$ in order to construct the bound, making them less useful in our context. The bound we will derive here, while correct, may be able to be improved.

Beginning with this definition of an unbiased estimator,
\begin{equation}
    \int \mathrm{d}\mathbf{x} \sin(\hat{\psi}(\mathbf{x}) - \psi) p(\mathbf{x}|\psi) = 0.
\end{equation}
We can take a derivative with respect to the parameter $\psi$:
\begin{equation}
    \int \mathrm{d}\mathbf{x} \sin (\hat{\psi}(\mathbf{x}) - \psi) \frac{\partial}{\partial \psi} p(\mathbf{x}|\psi) - \int \mathrm{d}\mathbf{x} p(\mathbf{x}|\psi) \cos (\hat{\psi}(\mathbf{x}) - \psi)= 0.
\end{equation}
The second term is just $-\langle \cos(\hat{\psi}-\psi)\rangle$. We can then follow a similar approach to the previous section, 
\begin{align}
\int \mathrm{d}\mathbf{x} \sin (\hat{\psi}(\mathbf{x}) - \psi) p(\mathbf{x}|\psi)\frac{\partial}{\partial \psi} \ln p(\mathbf{x}|\psi) &= \langle \cos(\hat{\psi}-\psi)\rangle, \\
\left[\int \mathrm{d}\mathbf{x} \sqrt{p(\mathbf{x}|\psi)} \sin (\hat{\psi}(\mathbf{x}) - \psi) \sqrt{p(\mathbf{x}|\psi)}\frac{\partial}{\partial \psi} \ln p(\mathbf{x}|\psi) \right]^2 &= \langle \cos(\hat{\psi}-\psi)\rangle^2. 
\end{align}
Applying Cauchy-Schwarz,
\begin{align}
    \langle \cos(\hat{\psi}-\psi)\rangle^2  &\leq \left[\int \mathrm{d}\mathbf{x} p(\mathbf{x}|\psi) \sin^2 (\hat{\psi}(\mathbf{x}) - \psi) \right] \left[ \int \mathrm{d}\mathbf{x} p(\mathbf{x}|\psi)\left(\frac{\partial}{\partial \psi} \ln p(\mathbf{x}|\psi)\right)^2 \right],\\
    \langle\cos(\hat{\psi}-\psi)\rangle^2 &\leq \langle \sin^2(\hat{\psi}-\psi) \rangle \mathcal{I}(\psi).
\end{align}
This bound is the analogous bound to the Cramer-Rao bound, but unfortunately this bound depends not only on the Fisher information $\mathcal{I}(\psi)$ but also $\langle \sin^2(\hat{\psi}-\psi) \rangle$ -- making it difficult to apply when we do not know the distribution of $\hat{\psi}$. We can get a more easily applied result -- at the cost of weakening the bound slightly. We start by rewriting $\langle \sin^2(\hat{\psi}-\psi) \rangle = 1 - \langle \cos^2(\hat{\psi}-\psi)\rangle$. Then 
\begin{equation}
        \langle\cos(\hat{\psi}-\psi)\rangle^2 \leq \left(1-\langle \cos^2(\hat{\psi}-\psi) \rangle\right) \mathcal{I}(\psi).
\end{equation}
Because $\langle \cos^2 (\hat{\psi}-\psi) \rangle \geq \langle \cos (\hat{\psi}-\psi) \rangle^2$ (by the positivity of the variance, or the Cauchy-Schwarz inequality again),
\begin{align}
            \langle\cos(\hat{\psi}-\psi)\rangle^2 &\leq \left(1-\langle \cos(\hat{\psi}-\psi) \rangle^2\right) \mathcal{I}(\psi), \\
            \left[1+\mathcal{I}(\psi)\right]\langle\cos(\hat{\psi}-\psi)\rangle^2 &\leq \mathcal{I}(\psi), \\
            \langle\cos(\hat{\psi}-\psi)\rangle &\leq \sqrt{\frac{\mathcal{I}(\psi)}{1+\mathcal{I}(\psi)}}.
\end{align}
This is Eq. (8) in the main text. This reduces to the ordinary Cramer-Rao bound in the limit of large $\mathcal{I}(\psi)$, in which case we expect the distribution of $\hat{\psi}$ becomes closely localized to $\psi$, so $\langle\cos(\hat{\psi}-\psi)\rangle \approx 1- \frac{1}{2}\langle (\hat{\psi}-\psi)^2 \rangle$. Simultaneously, $\sqrt{\mathcal{I}/(1+\mathcal{I})} = \sqrt{1/(\mathcal{I}^{-1}+1)} \approx 1-\frac{1}{2\mathcal{I}}$, so the bound Eq. (8) becomes Eq. (7).

{\section{\label{app:types} Generalization to multiple sensor types}}

{We briefly mention here the possibility that there are multiple sensor types with different properties. This would seem reasonable, as any membrane-bound molecule with charge could serve as a sensor if the cell can interpret its location reliably. There are also known multiple receptor types for chemoattractants like cAMP \cite{chen1996signaling}. If each sensor has its own value of $\beta_i = \mu_i R / D_i$, then the probability density of a single sensor is $p_i(\theta) = Z^{-1} e^{\kappa_i \cos (\theta-\psi)}$ with $Z = 2 \pi I_0(\kappa_i)$ and $\kappa_i = \beta_i E$. Then the log-likelihood for $N$ sensors is
\begin{equation}
    \ln \mathcal{L}(\psi,E;\bm{\theta}) = - \sum_i \ln \left[2\pi I_0(\kappa_i)\right] +  \sum_{i=1}^N \kappa_i \cos (\theta_i - \psi).
\end{equation}
and the Fisher information for the angle is
\begin{align}
    \mathcal{I}(\psi) &= -\left\langle \frac{d^2}{d\psi^2} \ln \mathcal{L}(\psi,E;\bm{\theta}) \right\rangle = \sum_{i=1}^N \kappa_i \langle \cos (\theta_i - \psi) \rangle= \sum_{i=1}^N \kappa_i \dfrac{I_1(\kappa_i)}{I_0(\kappa_i)}.
\end{align}
Importantly, in the limit of weakly polarized cells, i.e. $\kappa_i \ll 1$ for all $i$, then we can expand the Bessel functions as before and find
\begin{equation}
\mathcal{I}(\psi) \approx \dfrac{1}{2}\sum_{i=1}^N \kappa_i^2 = \dfrac{E^2}{2} \sum_{i=1}^N \beta_i^2.
\end{equation}
This means that {\it even if there are multiple sensors}, we should still expect to see $\langle \cos (\hat{\psi}-\psi) \rangle \approx \sqrt{\gamma^2 E^2 / (1+\gamma^2 E^2)}$ if cells are near their optimal sensing abilities. However, in this case, $\gamma^2 = \sum_{i=1}^N \beta_i^2/2$ is an effective averaged value for the different sensors. 
}

{\section{\label{app:downstream}Effects of downstream noise and alternate models for neural crest data}
\begin{figure}[htb]
    \centering
    \includegraphics[width=0.6\textwidth]{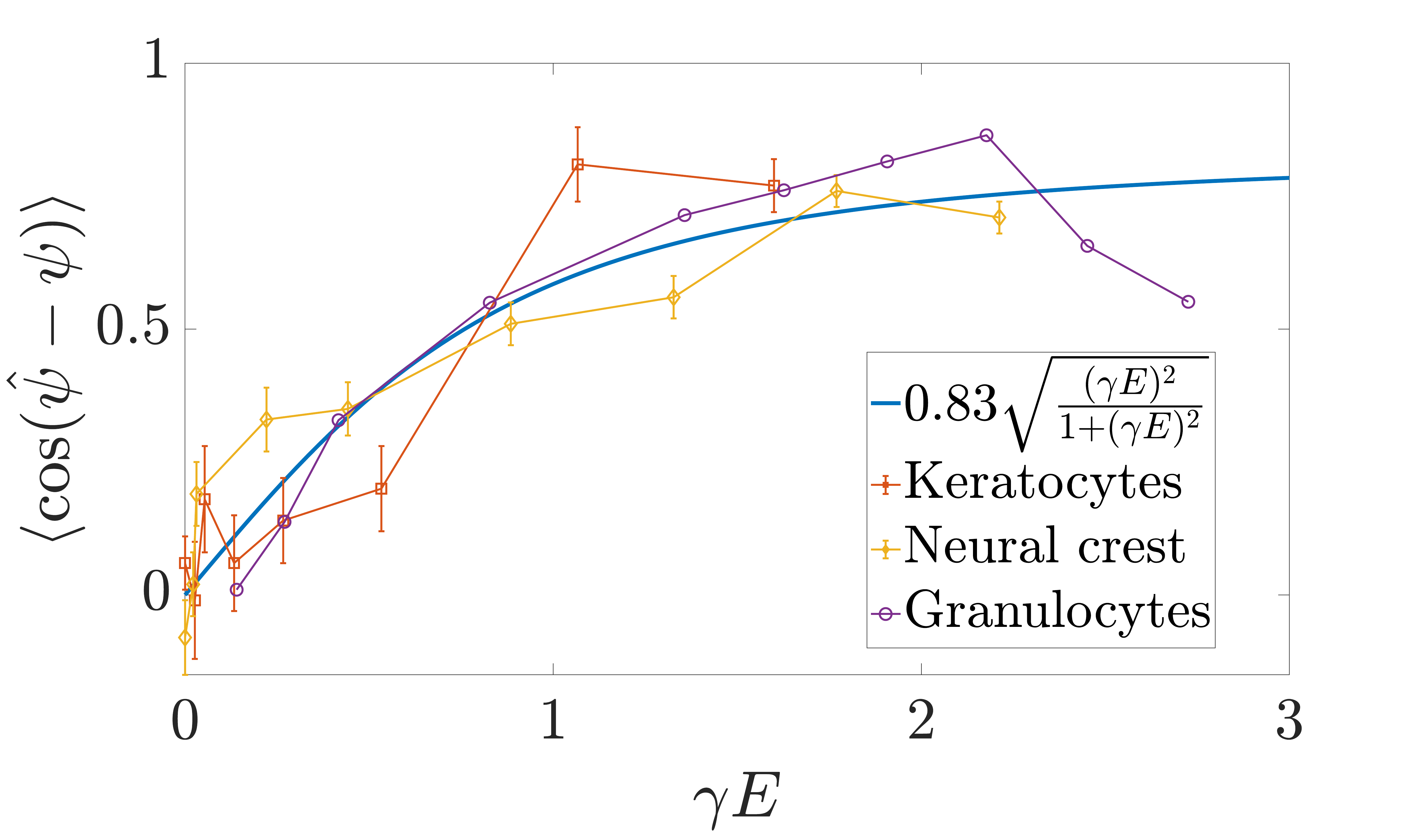}
    \caption{{Alternative fit with model using a maximum value of $\sim$ 0.83.}}
    \label{fig:fitalternative}
\end{figure}
In the main text, we have essentially assumed that single cells perfectly follow their estimation of the field direction $\hat{\psi}$. However, we know that there are other sources of noise in directed cell migration, both in processing the signal \cite{ten2016fundamental} and in stochastic events in motility itself. This will be particularly important at large field strengths, where the cell gets a large amount of information from the field -- the downstream noise can be the dominant source of error, as has been observed in chemotaxis at large gradient strengths \cite{fuller2010external}. To model this downstream noise, we assume that the cell, instead of following the estimator $\hat{\psi}$ perfectly, instead travels in a direction $\psi_v  = \hat{\psi} + \delta$, where $\delta$ is a random noise that is symmetric around $\delta = 0$ and independent from the sensor locations. %
If this is the case, then the directionality of the cell's velocity $\psi_v$ from the true field direction is reduced from our bound,
\begin{align}
    \langle\cos{(\psi_v-\psi)}\rangle &= \langle\cos{(\hat{\psi}-\psi+\delta)}\rangle \\
 &=\langle\cos{(\hat{\psi}-\psi})\cos\delta-\sin{(\hat{\psi}-\psi})\sin\delta\rangle\\
&=\langle\cos{(\hat{\psi}-\psi})\rangle\langle\cos\delta\rangle-\langle\sin{(\hat{\psi}-\psi})\rangle\langle\sin\delta\rangle \\
&= \langle\cos{(\hat{\psi}-\psi})\rangle\langle\cos\delta\rangle, \label{eq:cos_added_noise}
\end{align}
where the last step comes from the assumption that the distribution of noise $P(\delta)$ is symmetric about $\delta = 0$. If we have $\delta$ uniformly distributed over a range $\pm \Delta$, then $\langle \cos \delta \rangle = \sin\Delta/\Delta$.
We see from the data in Fig. 3 in the main paper that directionality, even at large fields, does not exceed a value of roughly $\sim 0.8$. We thus choose $\Delta=\pi/3$ or $\langle\cos\delta\rangle\approx0.83$ and fit the experimental data to Eq. \eqref{eq:cos_added_noise}, assuming that $\langle\cos{(\hat{\psi}-\psi})\rangle$ is given by the bound of Eq. (8) in the main text (Fig. \eqref{fig:fitalternative}). The fit quality is similar with this set of assumptions to that in Fig. 3. However, if we choose this alternate assumption, we find different, larger values for $\gamma$ relative to Fig. 3 in the main text ($\gamma$-values are $2.7\times10^{-3}$, $4.4\times10^{-3}$, and $2.7\times10^{-3}$ mm/mV for keratocytes, neural crest, and granulocytes, respectively). If we assume (motivated by measurements of the polarity cathode/anode ratio being 2.0) $\beta=3.5\times10^{-4}$ mm/mV as in the main text, the estimated number of sensors increases to $170-480$. If we assume a fixed number of 10$^4$ sensors, the ratio of cathode/anode concentration at 1000 mV/mm would be $1.10-1.17$, i.e. the level of polarization required to explain the data would increase by about 6\% from the previous maximum cathode/anode ratio of 1.10. The increase in required polarization or sensors when we assume downstream noise makes sense. Given the downstream noise, the directionality for a fixed amount of information decreases, so the cell would need more sensors or be more polarized (larger $\beta$) to achieve the same level of directionality as before. However, the assumption of downstream noise at these moderate levels does not qualitatively change the core predictions of the paper that cells may sense with small numbers (a few hundred) of sensors that are highly polarized or large numbers of weakly-polarized sensors.

\subsection{Neural crest data fit separately}
We note that in particular, the neural crest data in Fig. 3 is not a perfect fit with the curve. This may reflect, to some extent, systematic or random errors -- we view the overall fit of Fig. 3 as fairly rough. However, an alternate view of this data is that, because directionality is large even at the lowest fields (but presumably zero at zero field), the cells must have near-perfect information at the low fields. In this view, the deviation from perfect directionality is due to downstream noise, which could then vary with electric field. To some extent, an alternative model like this is difficult to disprove -- we can always invoke increasingly complicated downstream processes. To be truly convinced of the effect of sensor diffusive noise, we would need to know what the sensor molecule(s) are and their concentration. We focus on the first four neural crest data points, which show the largest deviation from our original model (Fig. \ref{fig:neuralcrest}). If we assume a downstream noise so $\langle\cos\delta\rangle\approx0.37$ ($\Delta=e-\pi/6$), we can fit the neural crest data separately. 
\begin{figure}[htb]
    \centering
    \includegraphics[width=0.6\textwidth]{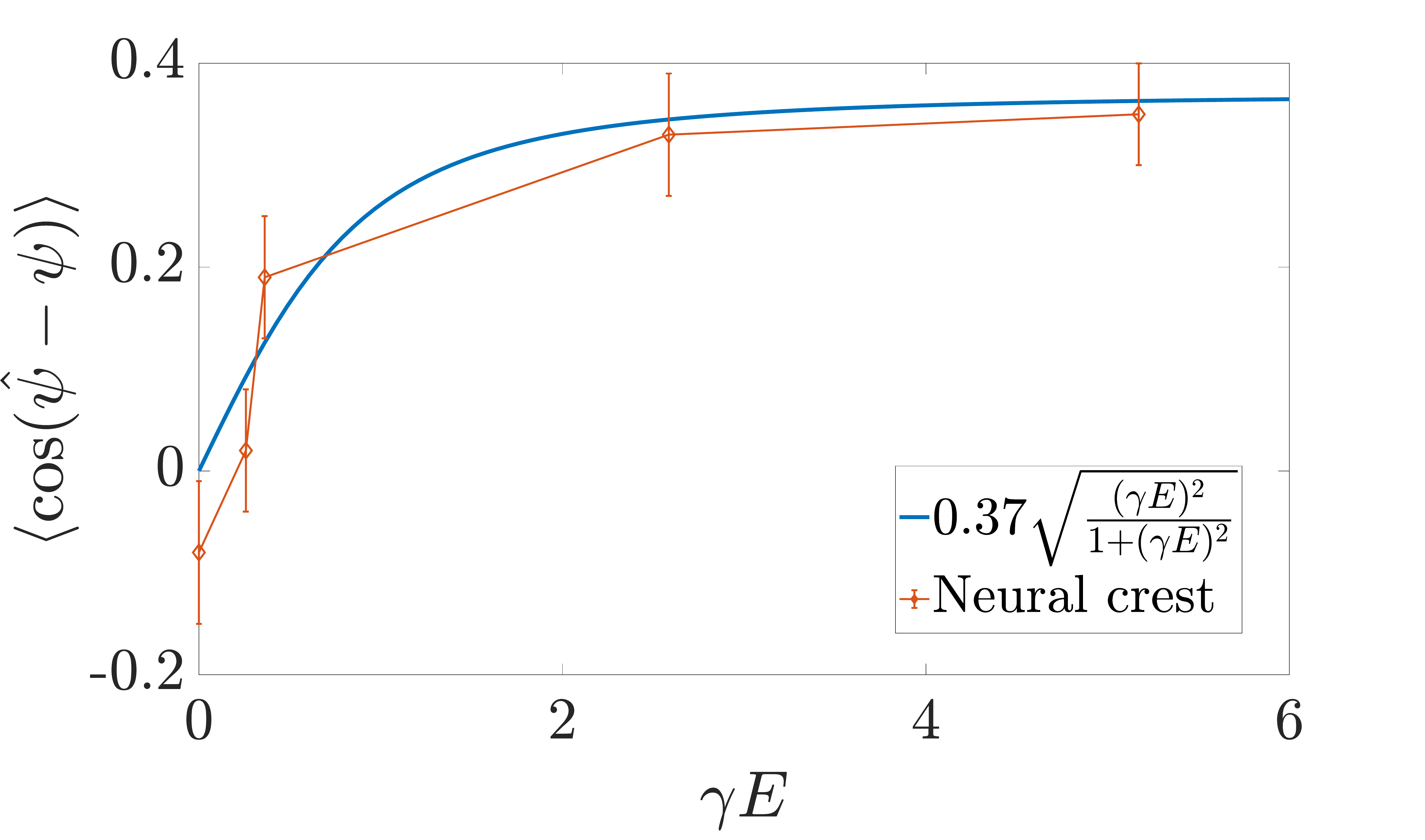}
    \caption{{First few neural crest data points fit to bound with downstream noise.}}
    \label{fig:neuralcrest}
\end{figure}
With this fit, we find a bigger $\gamma-$value of $5.2\times10^{-2}$ mm/mV. If we assume $\beta=3.5\times10^{-4}$ mm/mV, this would correspond to a number of sensors of about $6.5\times10^4$ -- about 50 sensors per square micron on a 10-micron-radius cell. Correspondingly, this $\gamma$ value would correspond to a cathode/anode ratio of $\sim 6$ if the number of sensors is fixed at 10$^4$. These values are larger than our predictions in the main text, but might also be plausible. Another possibility for the deviation between our model and the neural crest cell data is that neural crest cell galvanotaxis is qualitatively different in some way from keratocyte and granulocyte galvanotaxis. In fact, paper that originally measured neural crest galvanotaxis \cite{gruler1991neural} speculates that neural crest galvanotaxis and granulocyte galvanotaxis occur through different mechanisms. This is also supported by data showing that the response of neural crest cells exposed to a field being turned on can be complex, with polarization occurring in two phases \cite{nuccitelli1989extracellular}. This two-stage polarization would not be seen in our model. However, if, for instance, additional sensors are expressed on the surface in response to fields turning on, this sort of directionality change might occur. } %

{\section{\label{app:radius_moved} Prediction of size dependence does not depend on number of sensors}}

{In Fig. 4 in the main paper, we plot the directionality as a function of cell radius predicted by our bound. We did this by picking a value $N = 10^4$, but the results do not depend on this choice. We show this in Fig. \ref{fig:radius_circle}. We start with $\gamma$ fit for keratocytes from Fig. 3 in the main text, $\gamma = 0.002$ mm/mV. Since we cannot separate $\beta$ and $N$ in our fit, we show several values of $N$ and then pick the corresponding value of $\beta$ at the reference radius, $\beta_0 = \gamma \sqrt{d/N} = \mu R_0 /D$, where $R_0$ is the reference keratocyte radius. As expected, because the Fisher information only depends on $\gamma$ in the limit of smaller electric fields, these lines all collapse (Fig. \ref{fig:radius_circle}). Similarly, we can perform stochastic simulations with different numbers of sensors $N$ with varying $\beta_0$, and compute the sensor-direction-sum MLE direction, and see that the directionalities predicted from these simulations are also independent of the choice of $N$. However, because we are in the limit of Fig. 1 where there are deviations between the MLE estimator and the best-possible accuracy, these are systematically below the bound. However, the dependence on radius is similar. If we fit our data to the MLE instead of to the bound, we would expect similar values, but a difference of about a factor of $2$ in the predicted number of receptors.}

\begin{figure}[htb]
    \centering
    \includegraphics[width=0.7\textwidth]{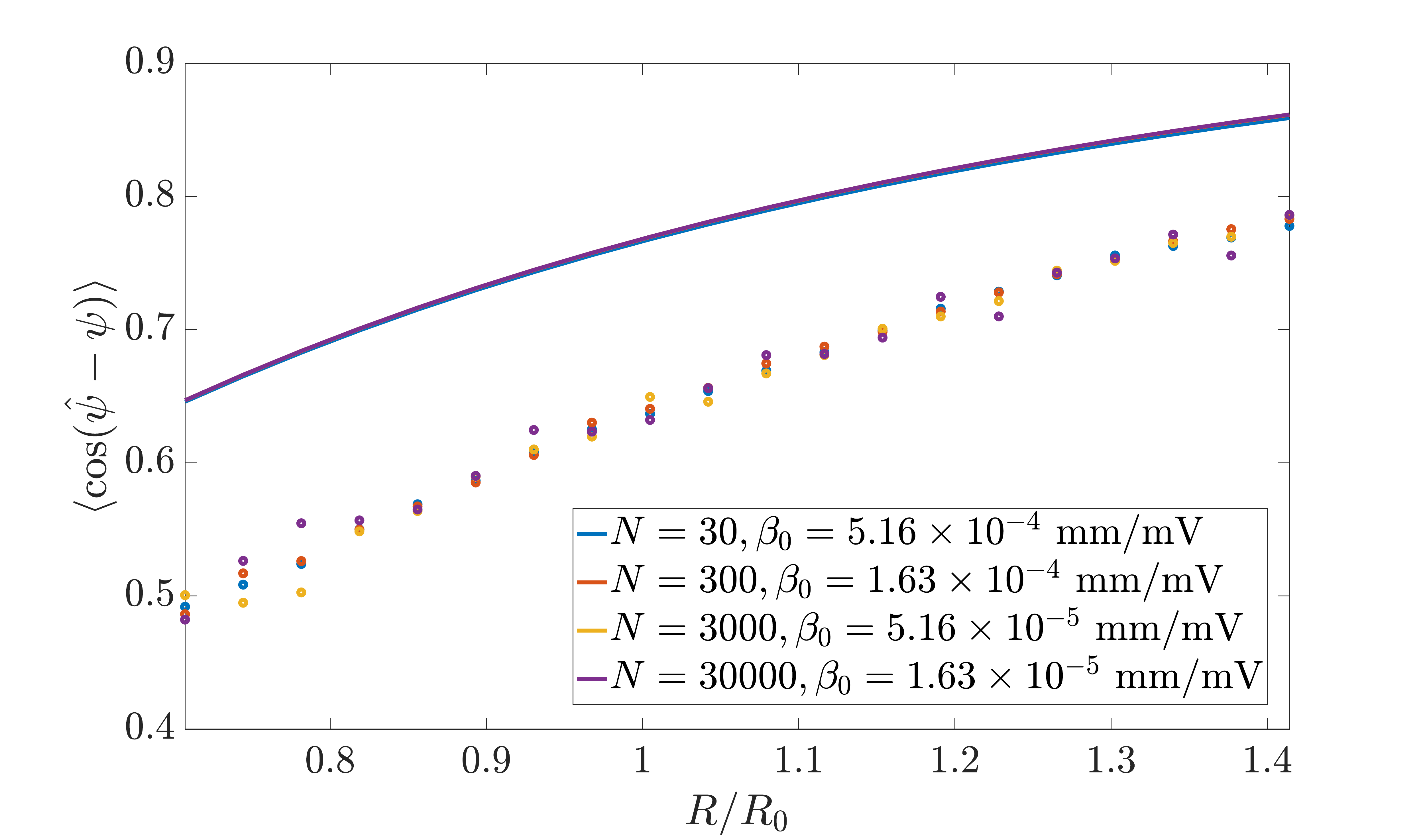}
    \caption{{Dependence of directionality on cell radius. Lines are the periodic bound with the Fisher information for a circle, symbols are stochastic simulation. $\gamma^2 = N\beta^2 / d$ is fit to the data of \cite{sun2013keratocyte}, $\gamma = 0.002$ mm/mV, and the E field used is 600 mV/mm. $d = 2$.}}
    \label{fig:radius_circle}
\end{figure}

The results on radius-dependence can also be generalized to the spherical cell assumption. {We see a near-identical dependence on radius from our model between sphere and circular cell} (Fig. \ref{fig:radius_sphere}). 
\begin{figure}[htb]
    \centering
    \includegraphics[width=0.7\textwidth]{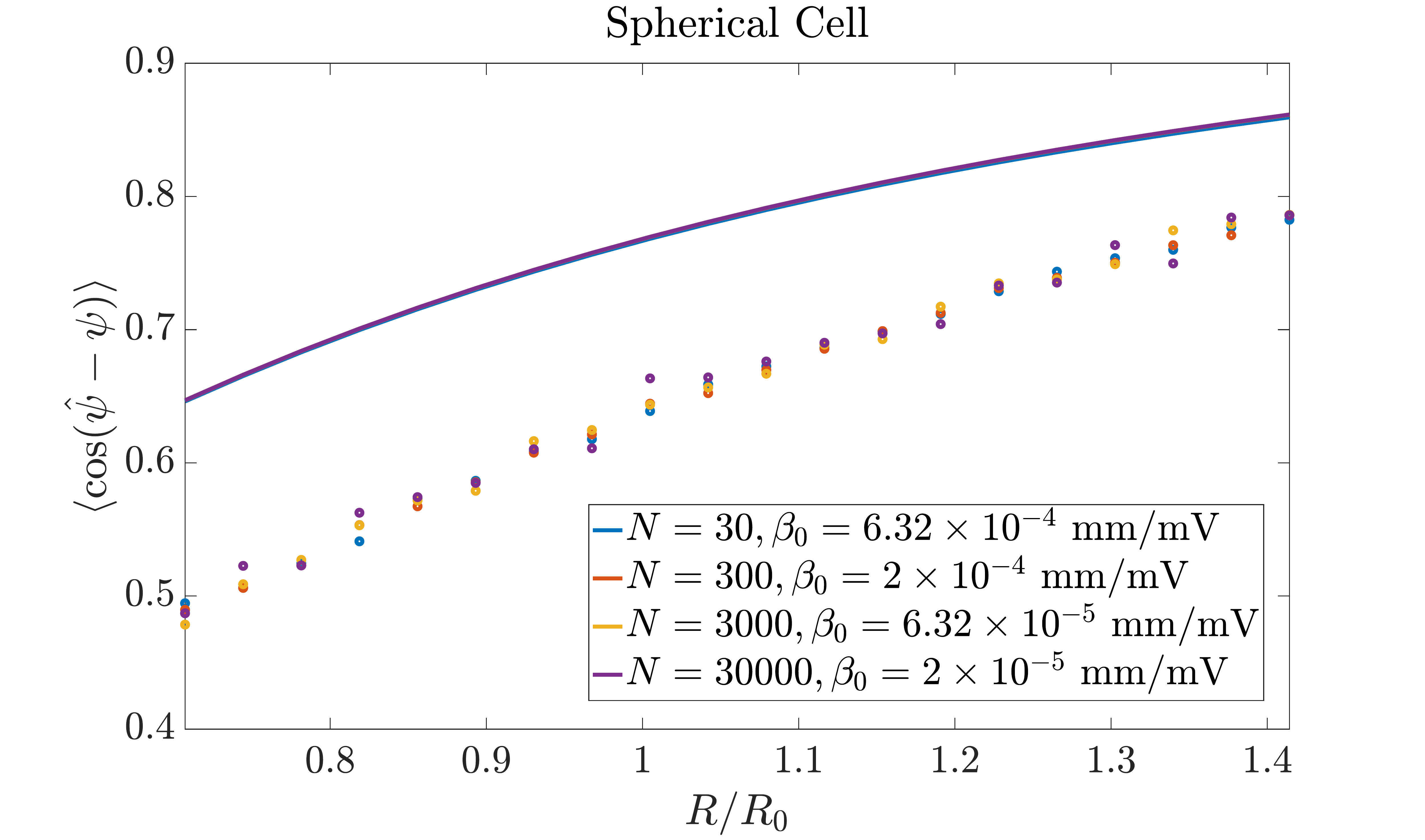}
    \caption{Dependence of directionality on cell radius assuming a spherical cell. Lines are the periodic bound with the Fisher information for a sphere, symbols are stochastic simulation. $\gamma^2 = N\beta^2 / d$ is fit to the data of \cite{sun2013keratocyte}, $\gamma = 0.002$ mm/mV, and the E field used is 600 mV/mm. $d = 3$.}
    \label{fig:radius_sphere}
\end{figure}
This is, of course, what we would expect {given Fig. 4 in the main text,} because the Fisher information $\mathcal{I}(\psi)$ differs only by a constant factor between $d = 2$ and $d = 3$, and this factor has been absorbed into $\gamma$. We can only extract $\gamma$ from the collapsed experimental data, but we also only need $\gamma$ to reliably predict the radius-dependence. %

{\section{\label{app:scaling} Alternate model: changing sensor number with cell size}}

{In the main text, we have assumed that, in comparing different cells with different radii, they still have the same number of sensors. This would be reasonable if, for instance, the cells have different areas because they have different levels of spreading on the surface -- e.g. if in three dimensions they have similar volumes and surface areas, but their measured areas when projected on the 2D substrate are different. However, if cells have different volumes due to, e.g. being in different stages in the cell cycle, we might expect different numbers of sensors. The most natural assumption would be then that the sensors are at a fixed {\it concentration} on the membrane, i.e. that the number per unit surface area of the cell is fixed (as a side note, some protein species tend to be kept at a fixed concentration, but others have concentrations that scale differently with cell size \cite{lanz2022increasing}; this is an active area of research). We show in Fig. \ref{fig:radius_nscale} how the results of Fig. 4 in the main text would differ if we instead assumed that sensor number is proportional to the cell surface area, e.g. choosing $N = (R/R_0)^2 N_0$. We see a stronger dependence on $R$, as we would expect, but again because of the relatively small range of radii and the large error bars, this is not very different in terms of comparison to experiment.}

\begin{figure}[htb]
    \centering
    \includegraphics[width=0.6\linewidth]{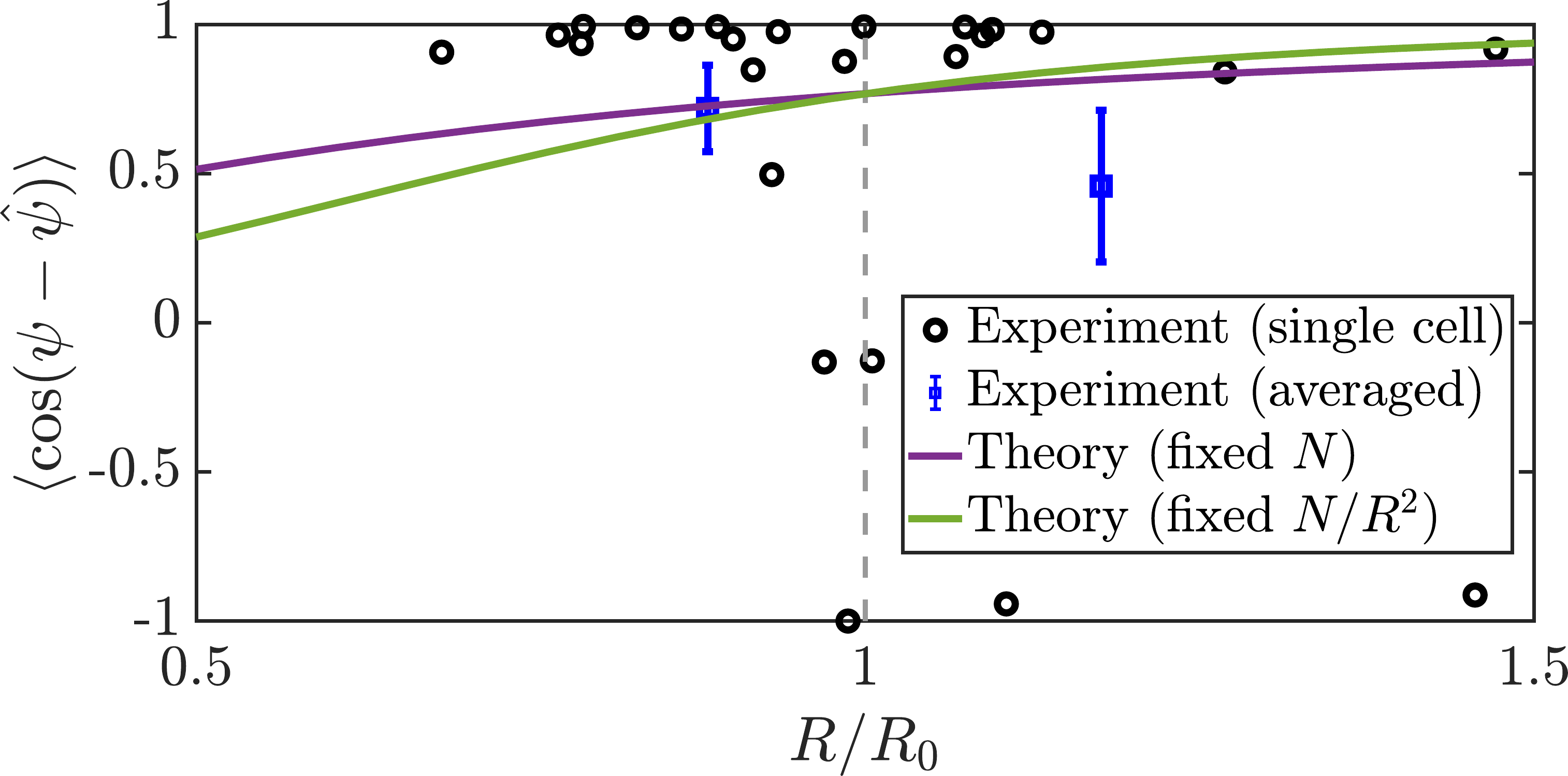}
    \caption{{Comparison between experiment and theory on cell size dependence using different assumptions about how $N$ scales with cell size. Lines are the periodic Cramer-Rao bound with the Fisher information for a circular cell, symbols are experimental data from \cite{sun2013keratocyte}. $\gamma = 0.002$ mm/mV is the fit to \cite{sun2013keratocyte} from Fig. 3 in the main text, and $E = 600$ mV/mm. $d = 2$.}}
    \label{fig:radius_nscale}
\end{figure}

{\section{\label{app:distributions_constant}Distribution of cell directionality under constant fields}}

{In Fig. \ref{fig:radius} in the main text, we see a broad distribution of directionalities from single cells, including directionalities near -1. This is a common feature of distributions of the cosine of an angle, and are also reproduced in our simulations. We show in Fig. \ref{fig:histogram_directionalitie_peaks} the distribution of directionality for a Brownian dynamics simulations with parameters appropriate to Fig. \ref{fig:radius}. We see that the distribution has peaks at both +1 and -1, and a broad population of cells with intermediate directionalities. This is not unique to galvanotaxis, as it is observed in many experiments on chemotaxis or other single-cell directed motility (for example, see Fig. 1c of \cite{skoge2014cellular}). }\\
\begin{figure}[htb]
    \centering
    \includegraphics[width=0.6\textwidth]{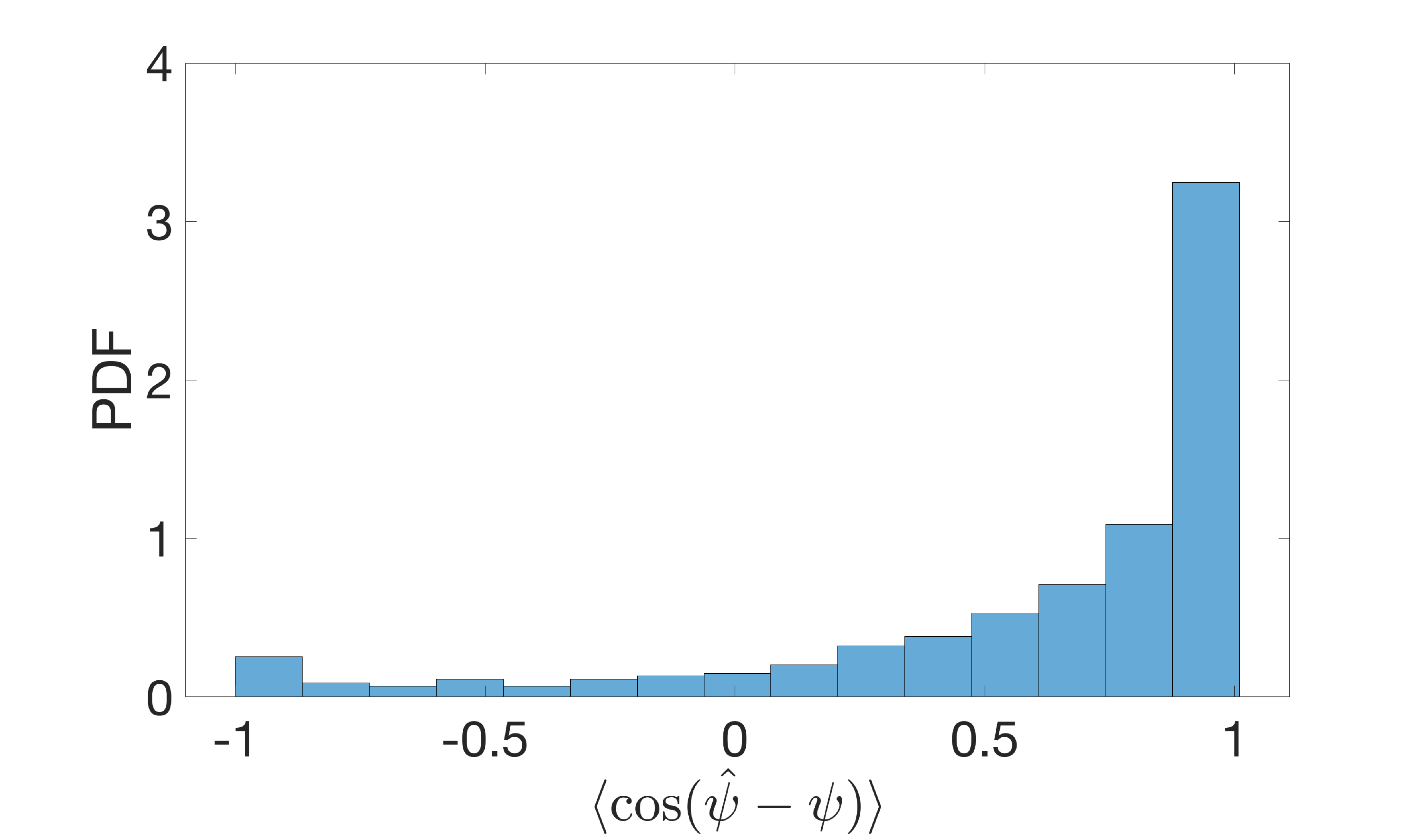}
    \caption{{Histogram of cell directionalities over simulation time course with constant electric field. $\kappa=0.0168$. $N=10000$ sensors. This corresponds to $\gamma=0.002$ mm/mV, to match the experiment in Fig. \ref{fig:radius}. $E=600$ mV/mm. This distribution was produced by first simulating 1000 cells with 10000 sensors using the Brownian Dynamics model Eq. \eqref{eq:langevin}, estimating $\hat{\psi}$ using Eq. \eqref{eq:receptorsum}, and then calculating the directionality.} }
    \label{fig:histogram_directionalitie_peaks}
\end{figure}

{Fundamentally, the broad distribution of directionalities in Fig. \ref{fig:histogram_directionalitie_peaks} and Fig. \ref{fig:radius} arises from changing variables from an angle $x$ to the cosine of the angle $\cos x$. Even if $x$ is uniformly distributed, the distribution  $P(\cos x)$ is broad and peaked at $\cos x = \pm 1$, because $\cos x$ is slowly varying when $\cos x \approx \pm 1$. This arises from a Jacobian factor in doing the change-of-variables for probability densities (see, e.g. Chapter 2 of \cite{kardar2007statistical}).}

\section{\label{app:switching}More details of cells in switching fields}
{We show a simulation using the same setup as Fig. \ref{fig:switching} but switch the field between larger angles, switching between $\pm\pi/2$ in Fig. \ref{fig:switching90}}.
\begin{figure}[htb]
    \centering
    \includegraphics[width=0.6\textwidth]{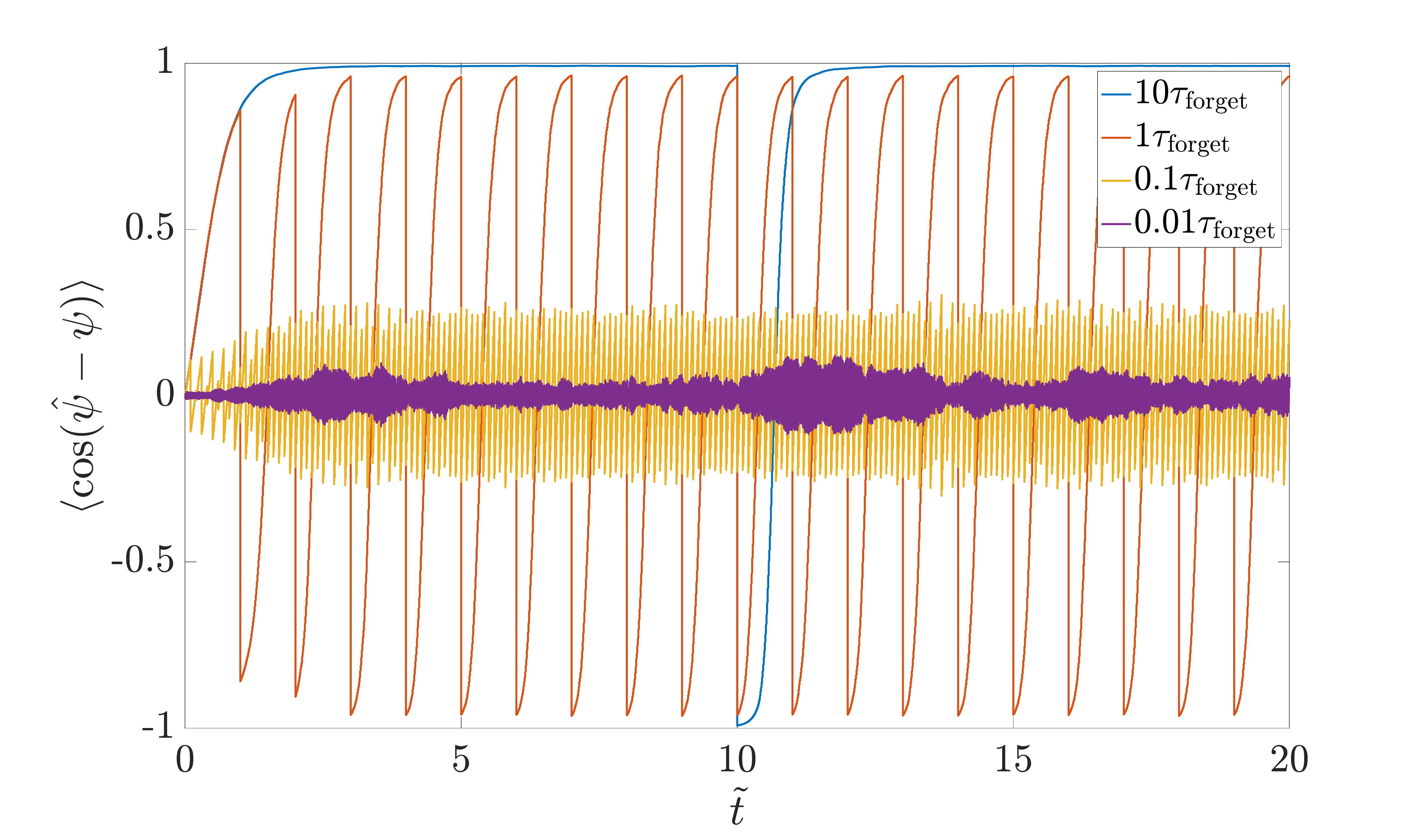}
    \caption{{{Directionality as a function of dimensionless time $\tilde{t}$ for switching angle $\pm\pi/2$ for different values of the exposure time ET. Parameters and simulation set up are the same as in Fig. 6b of the main text.}}}
    \label{fig:switching90}
\end{figure}

In the main text (Fig. 6a) we see the averaged directionality over 1000 cells as a function of time for different exposure times (ET) of a switching field. The distribution of the individual cell directionalities (Fig. \ref{fig:histogram_directionalities}) reveals that ET affects the distribution skewness. Lower exposure times have relatively small skew, which aligns with the rapid oscillation we see in Fig. 6a of the main text that are fairly symmetric. However, higher ET shows cells spending more and more time aligned with the field, skewing the directionalities closer to a value of one. 
\begin{figure}[htb]
    \centering
    \includegraphics[width=0.6\textwidth]{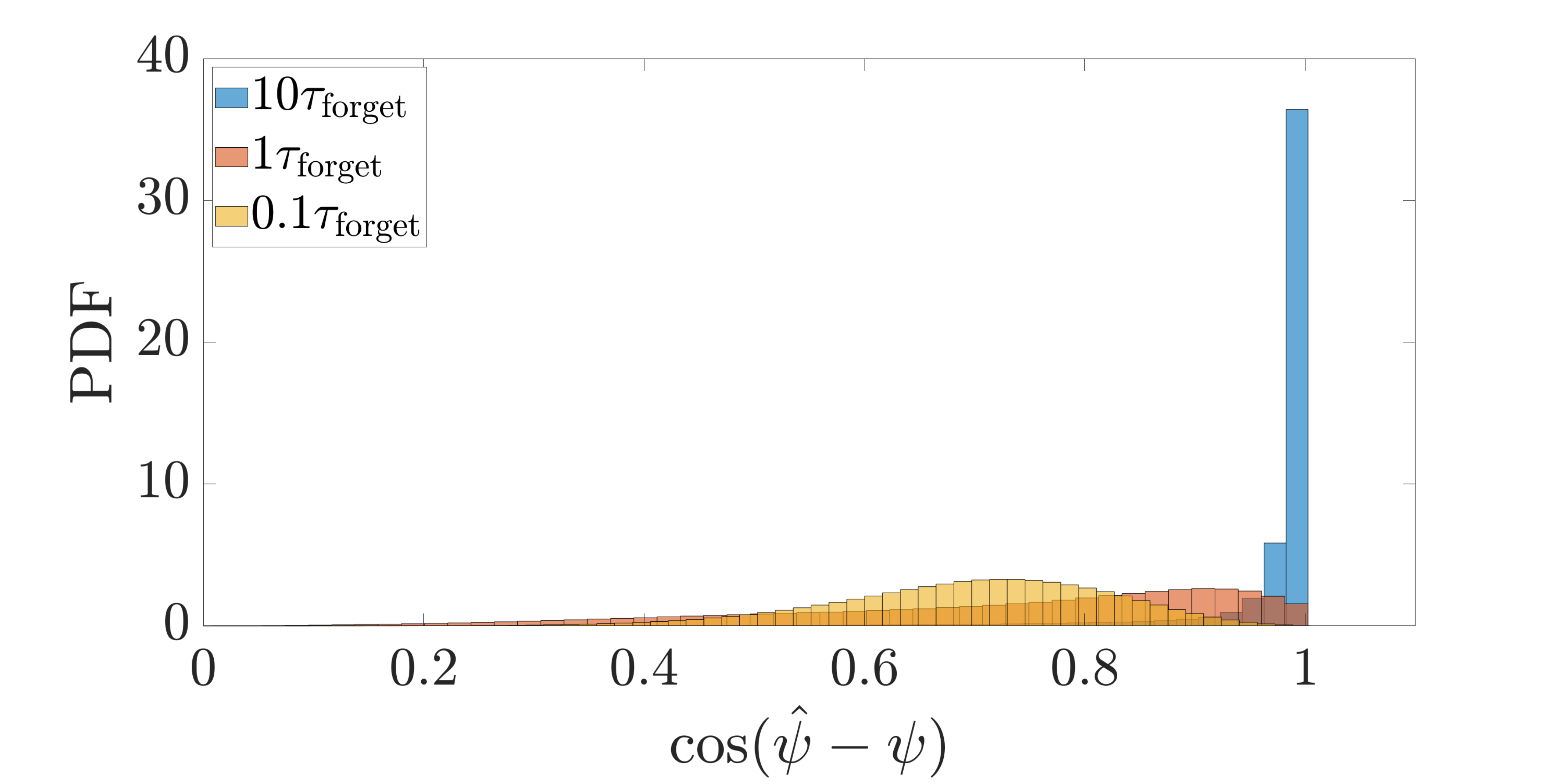}
    \caption{{Histograms of cell directionalities over simulation time course with switching electric field for different ET.}}
    \label{fig:histogram_directionalities}
\end{figure}

{\section{\label{app:protrusionmodel} Simple model to compute the maximum-likelihood estimator}}

{The core idea of this simple model is that the cell makes local protrusions normal to its boundary where there is a high concentration of local sensor. This is consistent with, e.g. recent work showing that local charge actually can regulate protrusion \cite{banerjee2022spatiotemporal}. The idea that cell direction and shape are controlled by protrusions normal to the boundary is a classic one \cite{lee1993principles,keren2008mechanism}. With this idea, we write a force density exerted by the cell at position $\theta$ as
\begin{equation}
    \mathbf{f}(\theta) = \hat{\mathbf{n}} \sum_{i=1}^N g(\theta - \theta_i),
\end{equation}
where $\hat{\mathbf{n}} = (\cos \theta,\sin \theta)$ is the local normal. Here, the sum $\sum_{i=1}^N g(\theta - \theta_i)$ is a way to create a smoothed picture of the local sensors -- if $g(\theta)$ were a delta function, this would be spikes at the location of sensors. We show a sketch of this function for a small number of sensors in Fig. \ref{fig:sensor_protrusion}.
\begin{figure}[htb]
    \centering
    \includegraphics[width=0.3\textwidth]{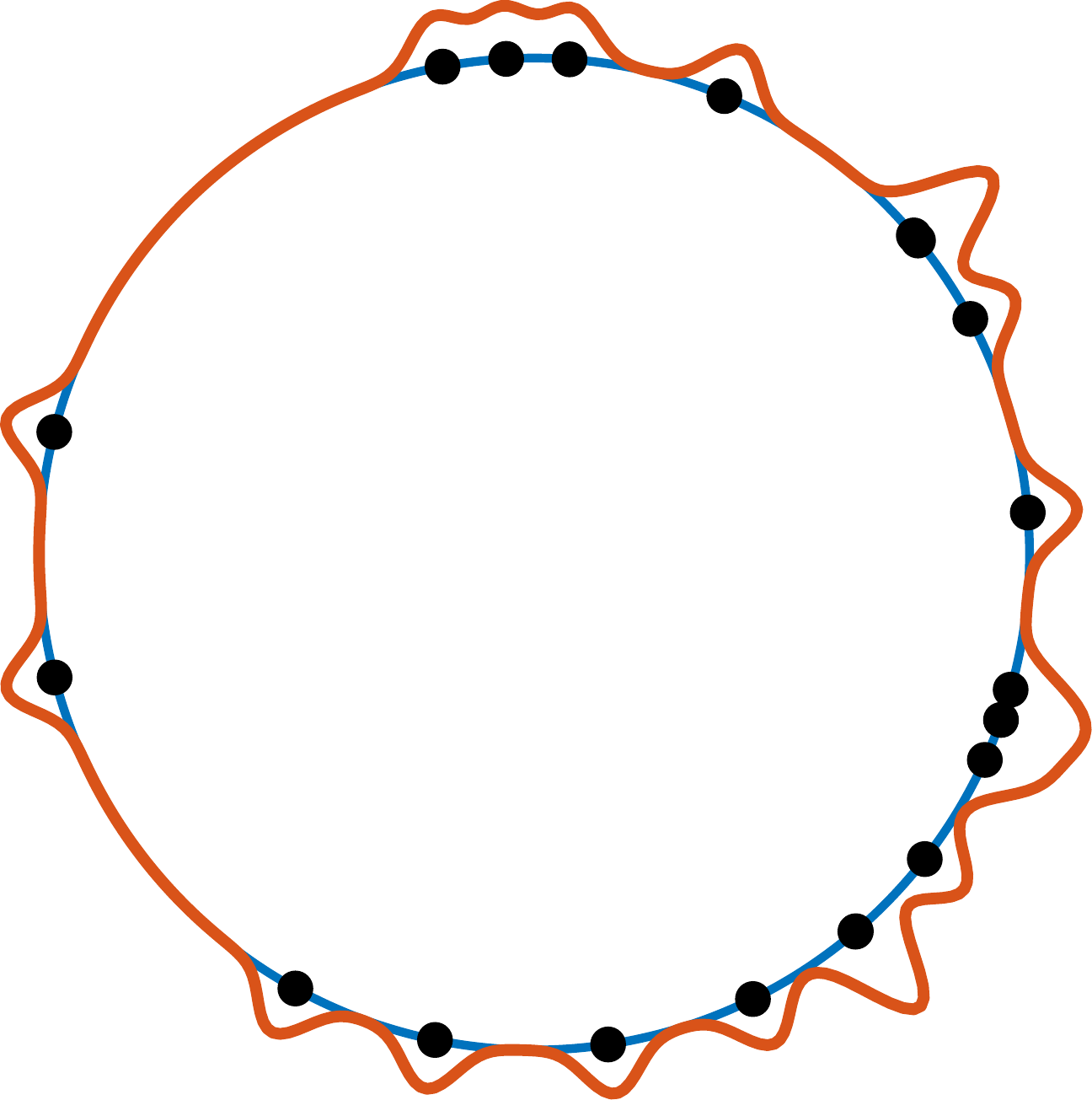}
    \caption{{Illustration of local protrusion model. Red line shows the sum $\sum_{i=1}^N g(\theta - \theta_i)$ for sensors (black dots).}}
    \label{fig:sensor_protrusion}
\end{figure}
The total force exerted by the cell's motility around the periphery is then just
\begin{equation}
\mathbf{F} = R\int \mathrm{d}\theta \mathbf{f}(\theta) = R \int \mathrm{d}\theta  \hat{\mathbf{n}} \sum_{i=1}^N g(\theta - \theta_i).
\end{equation}
We find that the direction of the force applied to the cell will be in the direction of the maximum-likelihood estimator of direction $\boldsymbol{\rho}$ as discussed in the main direction. This will happen with most reasonable functions $g(u)$, if we assume two very natural properties of the function $g(u)$. First, $g$ should be symmetric around $u = 0$ -- assuming otherwise would create a local chirality of the cell. Second, $g(u)$ should be $2\pi-$periodic, $g(u+2\pi) = g(u)$. The $x$ component of the force is
\begin{align}
F_x &= R \int_0^{2\pi} \mathrm{d}\theta \cos \theta \sum_{i=1}^N g(\theta - \theta_i) \\
    &= R\sum_{i=1}^N \int_0^{2\pi} \mathrm{d}\theta \cos \theta  g(\theta - \theta_i) \\
    &= R\sum_{i=1}^N \int_{-\theta_i}^{2\pi-\theta_i} \mathrm{d}u \cos (u+\theta_i)  g(u) \\
    &= R\sum_{i=1}^N \int_{-\theta_i}^{2\pi-\theta_i} \mathrm{d}u \left[\cos u \cos \theta_i -\sin u \sin \theta_i \right]  g(u),
\end{align}
where we have made the substitution $u = \theta - \theta_i$ and used the cosine angle addition formula. Then, we find
\begin{align}
F_x = R\sum_{i=1}^N \left[  \cos \theta_i \int_{-\theta_i}^{2\pi-\theta_i} \mathrm{d}u \cos u \, g(u) - \sin \theta_i \int_{-\theta_i}^{2\pi-\theta_i} \mathrm{d}u \sin u \, g(u) \right].
\end{align}
Assuming the periodicity of $g(u)$, the integrals over the region $[-\theta_i,0]$ would be exactly the same as integrating over the region $[2\pi-\theta_i,2\pi]$, so we get
\begin{align}
F_x &= R\sum_{i=1}^N \left[  \cos \theta_i \int_{0}^{2\pi} \mathrm{d}u \cos u g(u) - \sin \theta_i \int_{0}^{2\pi} \mathrm{d}u \sin u g(u) \right]\\
 &= R\sum_{i=1}^N \cos \theta_i \int_{0}^{2\pi} \mathrm{d}u \cos u g(u)\\
    &= F_0 \sum_{i=1}^N \cos \theta_i,
\end{align}
where we have noted that, because $g(u)$ is even and periodic, $\int_{0}^{2\pi} \mathrm{d}u \sin u g(u) = 0$, and $F_0 = R \int_{0}^{2\pi} \mathrm{d}u \cos u g(u)$ is a constant. Similarly, for the $y$ component of the force
\begin{align}
F_y &= R \int_0^{2\pi} \mathrm{d}\theta \sin \theta \sum_{i=1}^N g(\theta - \theta_i) \\
    &= R\sum_{i=1}^N \int_0^{2\pi} \mathrm{d}\theta \sin \theta  g(\theta - \theta_i) \\
    &= R\sum_{i=1}^N \int_{-\theta_i}^{2\pi-\theta_i} \mathrm{d}u \sin (u+\theta_i)  g(u) \\
    &= R\sum_{i=1}^N \int_{-\theta_i}^{2\pi-\theta_i} \mathrm{d}u \left[\sin u \cos \theta_i + \cos u \sin \theta_i \right]  g(u) \\
    &= R\sum_{i=1}^N \left[  \cos \theta_i \int_{-\theta_i}^{2\pi-\theta_i} \mathrm{d}u \sin u g(u) + \sin \theta_i \int_{-\theta_i}^{2\pi-\theta_i} \mathrm{d}u \cos u g(u) \right] \\
    &= R\sum_{i=1}^N \left[  \cos \theta_i \int_{0}^{2\pi} \mathrm{d}u \sin u g(u) + \sin \theta_i \int_{0}^{2\pi} \mathrm{d}u \cos u g(u) \right]\\
 &= R\sum_{i=1}^N \sin \theta_i \int_{0}^{2\pi} \mathrm{d}u \cos u g(u) \\
    &= F_0 \sum_{i=1}^N \sin \theta_i.
\end{align}
We thus see that the total force applied to the cell will be in exactly the direction $\boldsymbol{\rho} = \sum_i (\cos\theta_i,\sin \theta_i)$. This will then lead to a motion in the direction $\boldsymbol{\rho}$; the velocity could be found by balancing the total exerted force with the drag on the cell, e.g. as done in a related model for collective chemotaxis \cite{malet2015collective}. The only requirement is that the cell is able to make local protrusions in a normal direction. This model is -- naturally -- somewhat of an oversimplification, as we have not included any representations of cell polarity or protrusion dynamics beyond the simple function $g(u)$. However, we argue that this model shows the essentially plausibility that the cell can compute the direction $\bm{\rho}$. 
}

\end{document}